\documentclass[pdflatex,iicol,sn-basic]{sn-jnl}
\pdfoutput=1

\usepackage{amsmath, amssymb}
\usepackage{graphicx}
\usepackage{booktabs}
\usepackage[capitalise]{cleveref}
\AddToHook{env/appendices/begin}{%
  \crefalias{section}{appendix}%
  \crefalias{subsection}{subappendix}%
  \crefalias{subsubsection}{subsubappendix}}
\makeatletter
\AtBeginDocument{%
  \renewenvironment{appendices}{%
    \setcounter{equation}{0}%
    \begin{oldappendices}%
      \gdef\theequation{\@Alph\c@section\arabic{equation}}%
  }{\end{oldappendices}}}
\makeatother
\usepackage{tikz}
\usetikzlibrary{matrix}
\definecolor{orcidgreen}{HTML}{A6CE39}
\def\orcidlogo{\tikz[baseline=-0.6ex]{%
  \fill[orcidgreen] (0,0) circle (0.8ex);
  \node[text=white, inner sep=0pt, font=\sffamily\bfseries\fontsize{4.5pt}{4.5pt}\selectfont] at (0,0) {iD};}}
\makeatletter
\newcommand{\authororcid}[1]{\g@addto@macro\artauthors{\,\orcid{#1}}}
\makeatother
\usepackage{braket}
\usepackage{multirow}
\usepackage{xcolor}
\usepackage{colortbl}
\usepackage{placeins}  %
\makeatletter
\def\sectionfont{\reset@font\fontfamily{\rmdefault}\fontsize{12bp}{14bp}\bfseries\selectfont\raggedright\boldmath}%
\def\subsectionfont{\reset@font\fontfamily{\rmdefault}\fontsize{11bp}{13bp}\bfseries\selectfont\raggedright\boldmath}%
\renewcommand\section{\@startsection{section}{1}{\z@}%
  {-10pt \@plus -2pt \@minus -2pt}{5pt}{\sectionfont}}
\renewcommand\subsection{\@startsection{subsection}{2}{\z@}%
  {-8pt \@plus -2pt \@minus -2pt}{4pt}{\subsectionfont}}
\makeatother
\usepackage{subcaption}
\DeclareCaptionFont{sncap}{\fontsize{8}{9.5}\selectfont}
\DeclareCaptionLabelSeparator{snfig}{\hskip.7em}
\DeclareCaptionLabelSeparator{sntab}{\hskip2mm}
\usepackage{xspace}

\newcommand{\CC}{\mathbb{C}}
\newcommand{\RR}{\mathbb{R}}
\newcommand{\TT}{T}
\newcommand{\LL}{\mathcal{L}}
\newcommand{\MM}{\mathcal{M}}

\newcommand{\diag}{\operatorname{diag}}
\renewcommand{\skew}{\operatorname{skew}}
\newcommand{\topk}{\operatorname{top-k}}

\newcommand{\proj}{\operatorname{proj}}

\newcommand{\iconH}{\tikz[baseline=-0.55ex]{%
  \draw[black!50, line width=0.5pt] (-0.26,0) -- (0.26,0);
  \draw[line width=0.6pt, rounded corners=0.8pt, fill=white] (-0.13,-0.13) rectangle (0.13,0.13);
  \node at (0,0) {\tiny $H$};}}
\newcommand{\iconM}{\tikz[baseline=-0.55ex]{%
  \draw[black!50, line width=0.5pt] (-0.26,0.17) -- (0.26,0.17);
  \draw[black!50, line width=0.5pt] (-0.26,-0.17) -- (0.26,-0.17);
  \draw[line width=0.5pt] (0,0.17) -- (0,-0.17);
  \fill (0,0.17) circle (0.03);
  \fill (0,-0.17) circle (0.03);
  \draw[line width=0.6pt, rounded corners=0.8pt, fill=white] (-0.13,-0.12) rectangle (0.13,0.12);
  \node at (0,0) {\tiny $M$};}}
\newcommand{\iconMij}{\tikz[baseline=-0.55ex]{%
  \draw[black!50, line width=0.5pt] (-0.30,0.17) -- (0.30,0.17);
  \draw[black!50, line width=0.5pt] (-0.30,-0.17) -- (0.30,-0.17);
  \draw[line width=0.5pt] (0,0.17) -- (0,-0.17);
  \fill (0,0.17) circle (0.03);
  \fill (0,-0.17) circle (0.03);
  \draw[line width=0.6pt, rounded corners=0.8pt, fill=white] (-0.18,-0.12) rectangle (0.18,0.12);
  \node at (0,0) {\resizebox{0.30cm}{!}{$\boldsymbol{M_{i,j}}$}};}}
\newcommand{\iconCX}{\tikz[baseline=-0.55ex]{%
  \draw[black!50, line width=0.5pt] (-0.26,0.17) -- (0.26,0.17);
  \draw[black!50, line width=0.5pt] (-0.26,-0.17) -- (0.26,-0.17);
  \fill (0,0.17) circle (0.045);
  \draw[line width=0.5pt] (0,0.17) -- (0,-0.24);
  \draw[line width=0.5pt] (0,-0.17) circle (0.07);
  \draw[line width=0.5pt] (-0.07,-0.17) -- (0.07,-0.17);}}
\newcommand{\iconUfour}{\tikz[baseline=-0.55ex]{%
  \draw[black!50, line width=0.5pt] (-0.48,0.17) -- (0.48,0.17);
  \draw[black!50, line width=0.5pt] (-0.48,-0.17) -- (0.48,-0.17);
  \draw[line width=0.6pt, rounded corners=0.8pt, fill=white] (-0.31,-0.27) rectangle (0.31,0.27);
  \node at (0,0) {\tiny $U^{(4)}$};}}
\newcommand{\iconU}{\tikz[baseline=-0.55ex]{%
  \draw[black!50, line width=0.5pt] (-0.26,0) -- (0.26,0);
  \draw[line width=0.6pt, rounded corners=0.8pt, fill=white] (-0.13,-0.13) rectangle (0.13,0.13);
  \node at (0,0) {\tiny $U$};}}
\newcommand{\iconX}{\tikz[baseline=-0.55ex]{%
  \draw[line width=0.6pt, rounded corners=1pt, fill=white] (-0.14,-0.20) rectangle (0.10,0.20);
  \node at (-0.02,0) {\tiny $\mathbf{x}$};
  \foreach \y in {0.12,0,-0.12} \draw[black!50, line width=0.5pt] (0.10,\y) -- (0.40,\y);}}
\newcommand{\iconFN}{\tikz[baseline=-0.55ex]{%
  \foreach \y in {0.12,0,-0.12} \draw[black!50, line width=0.5pt] (-0.31,\y) -- (-0.13,\y);
  \draw[line width=0.6pt, rounded corners=1pt, fill=white] (-0.13,-0.19) rectangle (0.27,0.19);
  \node at (0.07,0) {\tiny $F_{N}$};
  \foreach \y in {0.12,0,-0.12} \draw[black!50, line width=0.5pt] (0.27,\y) -- (0.45,\y);}}

\newcommand{\iconT}{\tikz[baseline=-0.55ex]{%
  \foreach \y in {0.12,0,-0.12} \draw[black!50, line width=0.5pt] (-0.31,\y) -- (-0.13,\y);
  \draw[line width=0.6pt, rounded corners=1pt, fill=white] (-0.13,-0.19) rectangle (0.39,0.19);
  \node at (0.13,0) {\tiny $\TT(\theta)$};
  \foreach \y in {0.12,0,-0.12} \draw[black!50, line width=0.5pt] (0.39,\y) -- (0.57,\y);}}
\newcommand{\iconRy}{\tikz[baseline=-0.55ex]{%
  \draw[black!50, line width=0.5pt] (-0.28,0) -- (0.28,0);
  \draw[line width=0.6pt, rounded corners=0.8pt, fill=white] (-0.16,-0.13) rectangle (0.16,0.13);
  \node at (0,0) {\tiny $R_y$};}}
\newcommand{\iconRyH}{\tikz[baseline=-0.55ex]{%
  \draw[black!50, line width=0.5pt] (-0.38,0) -- (0.38,0);
  \draw[line width=0.6pt, rounded corners=0.8pt, fill=white] (-0.26,-0.17) rectangle (0.26,0.17);
  \node at (0,0) {\tiny $R_y^H$};}}
\newcommand{\iconCRy}{\tikz[baseline=-0.55ex]{%
  \draw[black!50, line width=0.5pt] (-0.28,0.17) -- (0.28,0.17);
  \draw[black!50, line width=0.5pt] (-0.28,-0.17) -- (0.28,-0.17);
  \draw[line width=0.5pt] (0,-0.17) -- (0,0.17);
  \fill (0,-0.17) circle (0.045);
  \draw[line width=0.6pt, rounded corners=0.8pt, fill=white] (-0.16,0.05) rectangle (0.16,0.29);
  \node at (0,0.17) {\tiny $R_y$};}}
\newcommand{\iconZ}{\tikz[baseline=-0.55ex]{%
  \draw[black!50, line width=0.5pt] (-0.26,0.17) -- (0.26,0.17);
  \draw[black!50, line width=0.5pt] (-0.26,-0.17) -- (0.26,-0.17);
  \draw[line width=0.5pt] (0,0.17) -- (0,-0.17);
  \fill (0,0.17) circle (0.045);
  \fill (0,-0.17) circle (0.045);}}

\begin{document}

\title[Fast Trainable Multilinear Bases]{Fast Trainable Multilinear Bases for Image Compression}

\author*[1]{\fnm{Shiwen} \sur{An}}\email{an.s.0e57@m.isct.ac.jp}
\authororcid{https://orcid.org/0000-0002-4727-2906}
\author[2]{\fnm{Zhongyi} \sur{Ni}}
\authororcid{https://orcid.org/0000-0002-6332-1237}
\author[2]{\fnm{Huanhai} \sur{Zhou}}
\author[2]{\fnm{Jin-Guo} \sur{Liu}}
\authororcid{https://orcid.org/0000-0003-1635-2679}

\affil[1]{\orgdiv{Department of Information and Communication Engineering}, \orgname{Institute of Science Tokyo}, \orgaddress{\city{Yokohama}, \postcode{226-8501}, \country{Japan}}}
\affil[2]{\orgdiv{Thrust of Advanced Materials}, \orgname{The Hong Kong University of Science and Technology (Guangzhou)}, \orgaddress{\city{Guangzhou}, \state{Guangdong}, \postcode{511453}, \country{China}}}

\abstract{\unboldmath The Discrete Fourier Transform (DFT), the Discrete Cosine Transform (DCT), and their block-wise variants underpin most deployed image and video codecs. Their effectiveness rests on three properties: their runtime is near-linear (up to a polylogarithmic factor) in the image size, they are exactly invertible, and they carry few to no parameters.
We generalize these transforms to dataset-adaptive isometric multilinear bases that keep all three properties at the cost of a polylogarithmic number of parameters, by parameterizing each basis as an isometric tensor network, inspired by quantum many-body theory, and training one shared basis per image dataset with Riemannian optimization.
Because the family contains the DFT and the DCT-IV (a variant of the DCT) as exact points, training can start from these transforms, and it improves on its starting point at the compression level it is trained for.
We evaluate the bases on natural photographs and line drawings. On photographs, which the cosine basis already fits well, the best trained basis leads the DCT only under aggressive truncation. On Quick Draw line drawings, under a common codec at matched quality, the best trained basis stores each image in about $20\%$ fewer bytes than the $8\times8$ block cosine transform of the Joint Photographic Experts Group (JPEG) format, not counting the one-off storage of the basis.}


\maketitle

\section{Introduction}
\label{sec:introduction}

Transform coding sits at the core of most deployed image and video codecs, from the Joint Photographic Experts Group (JPEG) standard \citep{wallace1992jpeg} to High Efficiency Video Coding (HEVC) \citep{sullivan2012hevc}. A fixed, fast transform maps image blocks or prediction residuals to coefficients, most of the signal energy lands in a few of them, and these coefficients are then quantized and entropy-coded \citep{goyal2001transform, mallat2008wavelet}. How well the transform concentrates energy sets how much the later stages can compress, so the transform is an important design choice in the pipeline. For decades this role has been filled by fixed trigonometric transforms. The most prominent are the Discrete Cosine Transform (DCT) \citep{ahmed1974dct} and its block-wise variants, together with the closely related Discrete Fourier Transform (DFT), whose fast algorithm, the Fast Fourier Transform (FFT) \citep{cooley1965fft}, supplies the template for computing such transforms quickly.

No fixed basis, however, serves every image dataset equally well. A fixed basis has no parameters, so it cannot adapt to any of them. For instance, the DCT comes close to the optimal Karhunen--Lo\`eve transform for sources near the first-order autoregressive (AR(1)) Gaussian regime \citep{jain1979sinusoidal}. Natural photographs sit near this regime but line drawings do not. To adapt the basis to the data, methods such as deep autoencoders \citep{theis2017, balle2017end} use trainable nonlinear encoding and decoding maps. These learned transforms usually compress better, but cannot take over the role of the traditional transforms, because the maps themselves carry far more parameters than a single image. A useful \emph{sparse basis}, one whose coefficients concentrate an image's energy in a few entries, should keep the properties that make the DCT and the FFT so effective:
\begin{enumerate}
    \item \textbf{Negligible parameter overhead}: the basis is described by a number of parameters that is negligible compared with the image size.
    \item \textbf{Fast to apply}: the transform and its inverse run in linear time, up to a factor polylogarithmic in the image size.
    \item \textbf{Exactly invertible}: the inverse map is explicit, so the transform is lossless before truncation.
\end{enumerate}
Under these constraints, we seek a systematic scheme that learns a basis for a given dataset, so that the dataset as a whole shrinks at the cost of storing only a negligible number of basis parameters.

Our insight comes from quantum computing, which supplies a precise dictionary between images and quantum states. An image with $N = 2^{n}$ pixels, flattened into a length-$N$ vector, has exactly the shape of an $n$-qubit state vector. A quantum circuit composes unitary gates into a reversible linear map, so from the image's point of view a circuit is simply a change of basis. Simulating a circuit of ${\rm poly}(\log N)$ gates exactly takes $O(N\,{\rm poly}(\log N))$ time on a classical device, near-linear in the image size. The FFT itself is such a circuit, the quantum Fourier transform (QFT) \citep{coppersmith1994approximate}, with its application cost derived in \cref{app:transform_complexity}. This dictionary has precedent in transform design: discrete wavelet transforms have been written as circuits of local unitary gates and new wavelets designed within that parameterization \citep{evenbly2018wavelets}, including wavelet families constructed with image compression in view \citep{mccord2022wavelets}. A quantum circuit is also mathematically equivalent to a tensor network \citep{markov2008simulating}, a multilinear map \citep{oseledets2011tensor} in which every tensor is constrained to preserve norms. For simplicity, we therefore parameterize the image basis as an isometric tensor network \citep{stoudenmire2016supervised, han2018unsupervised} rather than as a quantum circuit \citep{mcclean2016variational, cerezo2021variational, schuld2019quantum, benedetti2019parameterized}.

What remains open is which network topology (the connectivity pattern of the tensors) best fits a given class of images, how to train it, and how the trained basis depends on the dataset. Since finding the best tensor decomposition is NP-hard even for a fixed structure \citep{hillar2013most}, as is training the parameters of a fixed circuit \citep{bittel2021training}, we settle the first two questions empirically.

\begin{figure*}[!tbp]
  \centering
  \includegraphics[width=\textwidth]{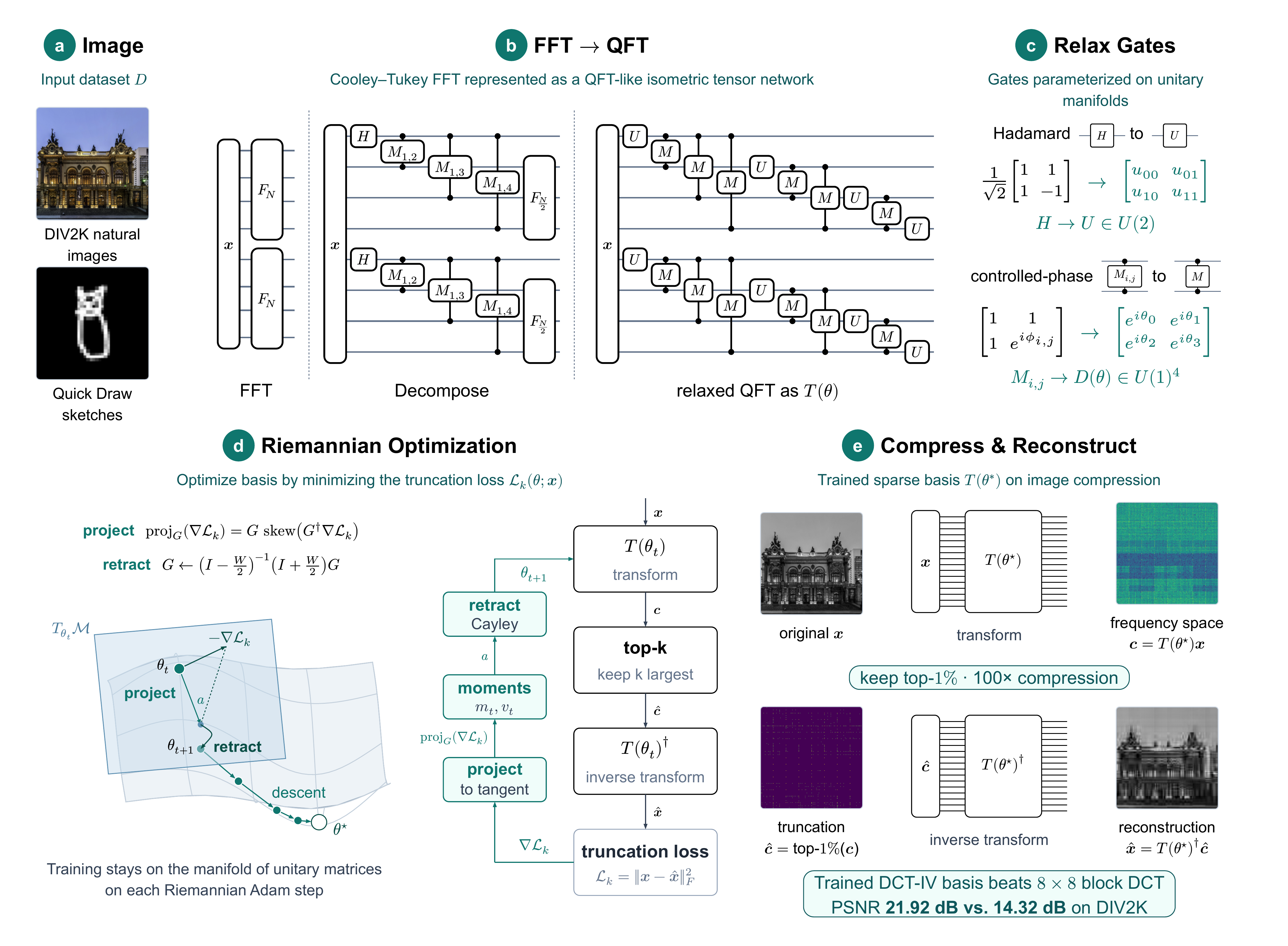}
  \caption{Overview of the framework. (a)~Training and evaluation datasets. The DIV2K example recurs in (e), and both examples are analyzed in \cref{sec:topology_div2k}. (b)~The Cooley--Tukey FFT decomposed into controlled phases $M_{i,j}$ and rewritten as the relaxed QFT network. (c)~Each gate family relaxed to a trainable unitary manifold. (d)~Riemannian training on the truncation loss: project the gradient onto the tangent space $T_{\theta_t}\mathcal{M}$, update the Adam moments $m_t, v_t$, and retract with the Cayley map. (e)~The trained basis applied at inference, here a DCT-IV keeping the top $1\%$ of coefficients, with mean DIV2K test peak signal-to-noise ratios (PSNR). Example images from DIV2K \citep{div2k} and Quick Draw (\citealp{quickdraw}; CC BY 4.0)}
  \label{fig:banner}
\end{figure*}

We construct trainable isometric tensor-network bases from the FFT, DCT-IV, and quantum-inspired variants (\cref{fig:banner}) and train them with Riemannian optimization \citep{becigneul2019riemannian, hauru2021riemannian}. For each topology, we learn \emph{one shared basis per dataset} from its training images and reuse it unchanged for every held-out image.
Training improves each basis over its initialization at the training keep ratio. On natural photographs (DIV2K; \citealp{div2k}), whose statistics the cosine basis already fits well, the best trained basis leads the DCT only under aggressive truncation. On non-photographic data such as Quick Draw line drawings \citep{quickdraw, ha2017neural}, the best trained basis stores each image in roughly $20\%$ fewer bytes than the DCT at the same reconstruction quality under a common codec, not counting the one-off storage of the basis.

The rest of the paper proceeds as follows. \Cref{sec:problem} formulates the per-dataset sparse basis problem, \cref{sec:circuits} constructs the isometric tensor-network topologies, \cref{sec:training} describes the Riemannian optimization, and \cref{sec:experiments} reports the main results.

\section{Problem formulation}
\label{sec:problem}

We seek a family $\mathcal{F}$ of unitary image bases satisfying the transform-coding constraints of \cref{sec:introduction}. Each basis $\TT(\theta) \in \mathcal{F}$ is represented by an isometric tensor network whose gate parameters $\theta$ lie on the product manifold $\MM$ of the gate factors (\cref{sec:alt_topologies}).

For a training sample $\mathcal{D} = \{\mathbf{x}^{(1)}, \dots, \mathbf{x}^{(L)}\}$ from a class of images and a retained-coefficient budget $k$, we learn one shared basis for the dataset by minimizing the mean post-truncation reconstruction error:
\begin{equation}
  \begin{aligned}
    \LL_k(\theta; \mathbf{x})
      &= \bigl\| \mathbf{x} - \TT(\theta)^{\dagger}\, \topk\!\bigl(\TT(\theta)\,\mathbf{x},\, k\bigr) \bigr\|_F^2, \\
    \theta^{\star}
      &= \operatorname*{arg\,min}_{\theta \in \MM}
        \; \frac{1}{L} \sum_{\ell=1}^{L}
        \LL_k\!\bigl(\theta; \mathbf{x}^{(\ell)}\bigr),
  \end{aligned}
  \label{eq:sparse_basis_problem}
\end{equation}
where $\lVert\cdot\rVert_F$ is the Frobenius norm and $\topk(\cdot,k)$ keeps the $k$ largest-magnitude coefficients and zeros the rest. After training, $\TT(\theta^\star)$ is stored once and applied to every held-out image from the dataset (\cref{fig:banner}e).

\begin{figure*}[!tbp]
\centering
\includegraphics[width=\textwidth]{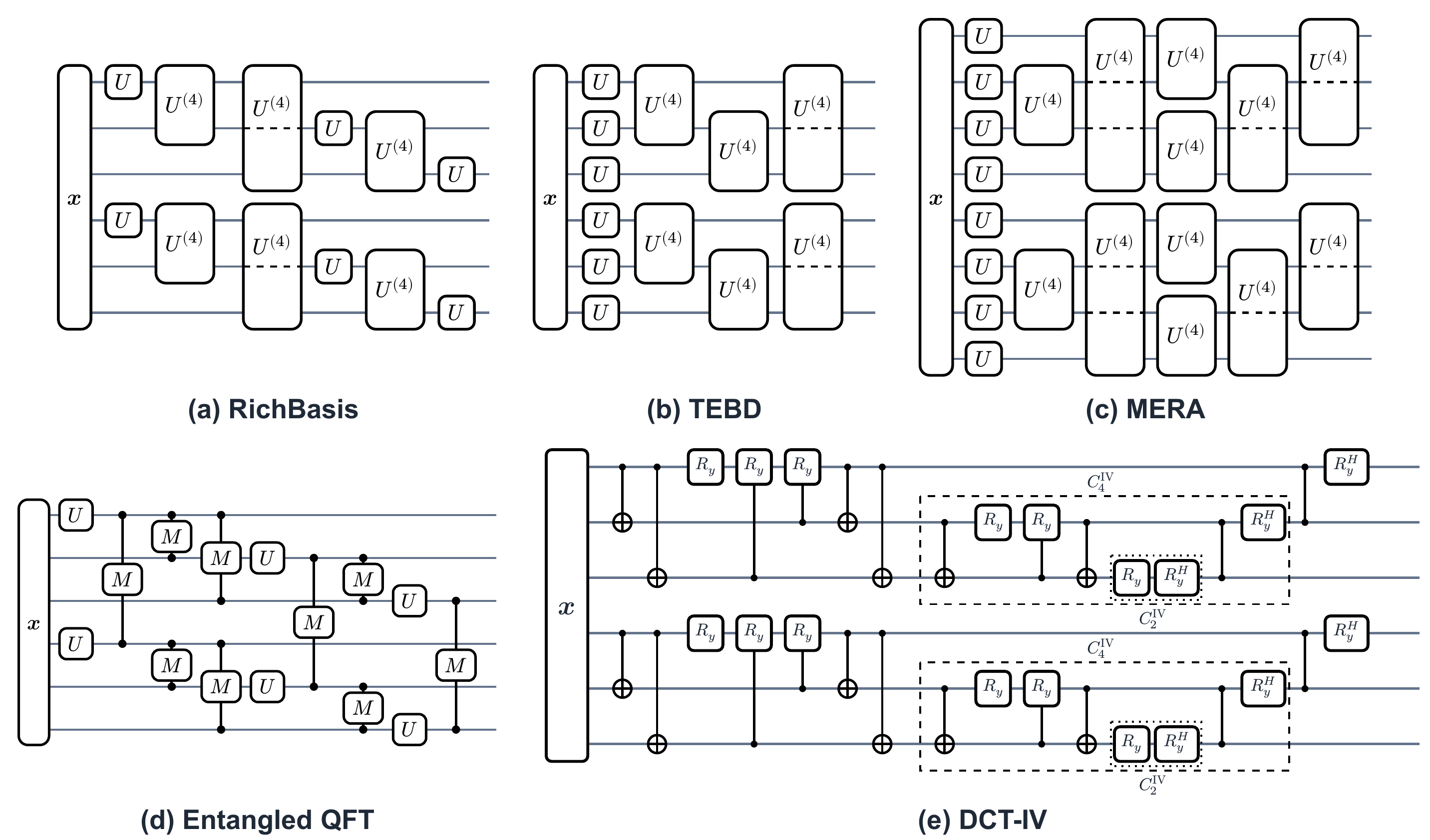}
\caption{Four circuit variants and the DCT-IV acting on input image $\mathbf{x}$, drawn at their relaxed gate families. (a)--(c) Topologies differing only in wiring, with relaxed Hadamard slots $U\in U(2)$ on one leg and general two-qubit gates $U^{(4)}\in U(4)$ on two (dashed wires pass through non-adjacent pairs). (d) Entanglement control based on existing QFT parametrization, within and across the two registers. Each $M$ is a relaxed controlled phase $D(\boldsymbol\theta)\in U(1)^4$ on its two dotted wires. (e) The relaxed DCT-IV decomposition, with merge gates $R_y^H$ relaxed on $O(2)$. Dots mark controls, $\oplus$ controlled-NOT targets, and dashed and dotted boxes the children $C_4^{\mathrm{IV}}$ and $C_2^{\mathrm{IV}}$}
\label{fig:topology_circuits}
\end{figure*}

\section{Isometric tensor-network bases}
\label{sec:circuits}

In order to find the best isometric tensor network $\TT(\theta)$'s structure, we realize a family $\mathcal{F}$ as tensor networks, whose small decomposed isometric tensors act directly on the binary digits of the image indices. An image $\mathbf{x} \in \RR^{2^m \times 2^n}$ has one row and one column index. Writing both indices in binary recasts $\mathbf{x}$ as a tensor with $m+n$ binary legs, drawn as one open wire per qubit,~\iconX{}. A basis $\TT(\theta)$~\iconT{} is then an ordered contraction of small isometric tensors, called gates as in a quantum circuit, each acting on one or two legs. Each gate $G$ is a $2\times2$ or $4\times4$ matrix on its legs, and being isometric it satisfies $G^\dagger G=I$, that is, it is unitary. Applying it amounts to attaching these gates to the image block~\iconX{} and absorbing them one at a time. The contraction order is fixed by the network's recursion rather than optimized; where the two-leg gates stay diagonal, fusing its levels gives the $O\!\left(2^{m+n}(m+n)\right)$ fast-transform schedule noted in \cref{app:gate_dictionary}. The construction of~\iconT{} begins with the classical FFT, which already has the network format.

\subsection{Construction from fast transforms}
\label{sec:qft}
For a $2^n \times 2^n$ image, the separable 2-D FFT maps $\mathbf{x} \mapsto F_N \mathbf{x} F_N^{\top}$, applying the per-axis DFT matrix $F_N$ of size $N=2^n$, computed by the Cooley--Tukey FFT, along each axis. The row and column indices of the image~\iconX{} thus form separate registers, each carrying a dense block $F_N$~\iconFN{} (\cref{fig:banner}b). One Cooley--Tukey step reduces $F_N$ to $F_{N/2}$: a Hadamard $H$~\iconH{} acts on the most-significant bit $i$, controlled phases $M_{i,j}$~\iconMij{} couple it to each remaining bit $j$, and the half-size transform continues on the remaining legs, with
\begin{align*}
  \iconH{} &= \tfrac{1}{\sqrt{2}}\bigl(\begin{smallmatrix}1&1\\1&-1\end{smallmatrix}\bigr), \\
  \iconMij{} &= M_{i,j} = \diag\bigl(1,1,1,e^{i\phi_{i,j}}\bigr),
\end{align*}
and the twiddle phase $\phi_{i,j}=2\pi/2^{\,j-i+1}$ fixed by the bit pair, as in the controlled $R_{j-i+1}$ gate of the standard QFT circuit. The recursion terminates at $F_2$, which is $H$ after the overall $N^{-1/2}$ normalization, and yields the QFT network, with one $H$ per wire and one $M_{i,j}$ per pair of bits. Contracting it with the image~\iconX{} reproduces the classical FFT up to a fixed bit reversal, the overall $N^{-1/2}$ normalization, and the conjugate-root convention; \cref{app:decompose} gives the full derivation.

We then make the tensor network $\TT$ trainable by fixing its wiring and relaxing each gate within its natural matrix manifold \citep{edelman1998geometry, absil2008optimization} (\cref{fig:banner}c); the last stage of \cref{fig:banner}b draws the resulting relaxed network. Each fixed Hadamard becomes an arbitrary unitary $U$, and each controlled phase keeps its diagonal form while freeing its diagonal entries into $D(\boldsymbol\theta)$:
\begin{align*}
  \iconH{}
  \;&\longrightarrow\; \iconU{} = U = \bigl(
  \begin{smallmatrix}u_{00} & u_{01} \\ u_{10}& u_{11}\end{smallmatrix}\bigr) \\
  \iconMij{} \;&\longrightarrow\; \iconM{} = D(\boldsymbol\theta) =
  \diag\bigl(e^{i\theta_0},e^{i\theta_1},e^{i\theta_2},e^{i\theta_3}\bigr),
\end{align*}
with every $\theta \in \mathbb{R}/2\pi\mathbb{Z}$, a relaxed Hadamard ranges over $U(2)$ and a relaxed phase over $U(1)^4$; \cref{app:gate_dictionary} tabulates the same fixed-to-relaxed map for every gate in the family.

The QFT topology contains $m+n$ Hadamard gates and $P=\frac{m(m-1)}{2}+\frac{n(n-1)}{2}$ compact phase gates, so its implemented search space is the product manifold
\begin{equation}
  \mathcal{M}_{\mathrm{QFT}}^{(m,n)}=U(2)^{m+n}\times\bigl(U(1)^4\bigr)^P,
  \label{eq:qft_product_manifold}
\end{equation}
with one factor, $U(2)$ or $U(1)^4$, for each gate. Every gate is therefore unitary, and the network applies the gates in sequence, each on its own legs, so $\TT_{\mathrm{QFT}}(\theta)\in U(2^{m+n})$ at every parameter setting and its inverse is the adjoint network, applied gate by gate in reverse order. The same argument holds for every topology in the family, and \cref{app:training_preservation} shows that training preserves this unitarity. The parametric QFT contains the classical FFT at the Fourier parameter values and supplies the template for the rest of the family.

The DCT-IV isometric tensor network follows the same recurse--decompose--relax process as the QFT. Its recursion produces the sparse real circuit in \cref{fig:topology_circuits}e; relaxing those gates within their manifolds preserves unitarity, with the rotation and mirror gates staying real orthogonal, and retains the exact DCT-IV as its initialization. \Cref{app:dct4_circuit} derives the recursion and bit reversal, with the gate-wise relaxation given in \cref{app:gate_dictionary}.

\begin{table*}[!tbp]
\centering
\caption{Properties of each topology for a one-dimensional transform of length $N=2^n$ on $n$ qubits: the wiring (\textit{Connectivity}), the numbers of \textit{Gates} and of trainable real \textit{Parameters}, the \textit{Depth} in sequential gate applications, and the \textit{Cost} of applying the transform. Ent.\ QFT, the Entangled QFT, counts its coupled register pair, and MERA requires a power-of-two $n$.}
\label{tab:circuits}
\tablebodyfont
\setlength{\tabcolsep}{4.5pt}
\begin{tabular*}{\textwidth}{@{\extracolsep{\fill}}lcccccc@{}}
\toprule
\textbf{Property} & \textbf{QFT} & \textbf{DCT-IV} & \textbf{Ent.\ QFT} & \textbf{TEBD} & \textbf{MERA} & \textbf{RichBasis} \\
\midrule
Connectivity & all-to-all & all-to-all & all-to-all + row--column & ring & hierarchical & all-to-all \\
Gates & $\tfrac{1}{2}n(n+1)$ & $\tfrac{3}{2}n(n-1) + 3n - 1$ & $n(n+2)$ & $2n$ & $3n - 2$ & $\tfrac{1}{2}n(n+1)$ \\
Parameters & $2n(n+1)$ & $\tfrac{13}{2}n(n-1) + 6n - 4$ & $4n(n+2)$ & $20n$ & $36n - 32$ & $4n(2n-1)$ \\
Depth & $O(\log^2 N)$ & $O(\log^2 N)$ & $O(\log^2 N)$ & $O(\log N)$ & $O(\log N)$ & $O(\log^2 N)$ \\
Cost & $O(N\log N)$ & $O(N\log^2 N)$ & $O(N^2 \log N)$ & $O(N\log N)$ & $O(N\log N)$ & $O(N\log^2 N)$ \\
\bottomrule
\end{tabular*}
\end{table*}

\subsection{Topology variants}
\label{sec:alt_topologies}
The remaining topologies vary the wiring and the local gate (\cref{fig:topology_circuits}a--d), following circuit patterns from quantum many-body physics. \emph{Entangled QFT} keeps the two QFT registers and adds one relaxed controlled phase $D(\boldsymbol\theta)$~\iconM{} between each matched row--column pair. Setting the phases in these entangled pair gates to zero recovers the separable QFT. \emph{TEBD} (time-evolving block decimation; \citealp{vidal2004tebd}), \emph{MERA} (multiscale entanglement renormalization ansatz; \citealp{vidal2007mera}), and \emph{RichBasis} share a single gate set, the $U(2)$ single-qubit layer with unrestricted two-qubit gates in $U(4)$, and differ only in how those gates are wired. \emph{TEBD} uses a nearest-neighbor ring, \emph{MERA} a hierarchy of disentangler and isometry layers, every gate here a square two-qubit unitary, and \emph{RichBasis} keeps the QFT's all-to-all wiring while relaxing each phase gate to a full two-qubit gate in $U(4)$.
Fixing the shape of $\TT(\theta)$ fixes the product manifold $\MM$ it optimizes at. \Cref{app:gate_dictionary} lists each reference gate beside the family it relaxes to and defines these manifolds rigorously in \cref{eq:appendix_gate_manifolds}.

Since each gate lies in a unitary or real orthogonal factor of $\MM$, every topology is unitary by the argument of \cref{sec:qft}, and its $O\!\left((m+n)^2\right)$ fixed-size gates (\cref{tab:circuits}, counted in \cref{app:transform_complexity}) keep the parameter overhead polylogarithmic in the image size, as \cref{sec:introduction} requires.

\section{Riemannian training}
\label{sec:training}

We use Riemannian optimization for the dataset-mean objective in \cref{eq:sparse_basis_problem}, and report the budget as the keep ratio $\rho = k/2^{m+n}$, the retained fraction of all coefficients. Each training step evaluates $\LL_k$ on a minibatch, averaged over its images, backpropagates it to a gradient on the gates (\cref{app:einsum}), and then updates each gate on its prescribed manifold. We hold out $15\%$ of the training slice for validation and return the parameters with the lowest validation loss.

\subsection{Truncation loss}
\label{sec:loss}
The loss has a coding interpretation that pins down what training can change. Write $\hat{\mathbf{x}}=\TT(\theta)^{\dagger}\topk\!\bigl(\TT(\theta)\,\mathbf{x},\,k\bigr)$ for the reconstruction drawn in \cref{fig:banner}d. Because $\TT(\theta)$ is unitary it preserves the Frobenius norm, so moving $\TT(\theta)$ inside the norm rewrites the pixel-domain error as the energy of the discarded coefficients, the sum of their squared magnitudes:
\begin{equation*}
  \begin{aligned}
    \LL_k
      &= \bigl\lVert \mathbf{x} - \hat{\mathbf{x}}\bigr\rVert_F^2 = \bigl\lVert \TT(\theta)\,\mathbf{x} - \topk\!\bigl(\TT(\theta)\,\mathbf{x},\,k\bigr)\bigr\rVert_F^2 .
  \end{aligned}
\end{equation*}
The same norm preservation fixes the total coefficient energy, $\lVert\TT(\theta)\,\mathbf{x}\rVert_F^2=\lVert\mathbf{x}\rVert_F^2$, for every $\theta$, so $\LL_k$ is the part of the image's energy that the $k$ retained coefficients fail to capture. For a fixed basis, keeping the $k$ largest-magnitude coefficients already retains as much energy as any choice of $k$ coefficients, so training can lower $\LL_k$ only by changing the basis. It rotates the gates inside $\TT(\theta)$ so that the energy of the training images concentrates on fewer coefficients, leaving less of it outside the retained set. The truncation itself stays outside the transform.

The backward pass differentiates through $\topk$, which is not smooth. It still yields a well-defined gradient because the selection it makes is piecewise constant in $\theta$: which $k$ coefficients are retained changes only when a discarded coefficient overtakes a retained one in magnitude. Between such crossings, $\topk$ applies one fixed selection, so the loss is simply the energy of a fixed set of discarded coefficients, a smooth function of $\theta$, and automatic differentiation returns its exact gradient $\nabla\LL_k$. A crossing is a parameter setting that a descent step essentially never lands on exactly, so the smoothed surrogates developed for discrete selections \citep{bengio2013estimating, jang2017categorical, xie2019reparameterizable} are not needed.

\subsection{Manifold-constrained update}
\label{sec:riemannian}

Every trainable gate $G$ of $\TT(\theta)$ must keep $G^\dagger G=I$ (\cref{sec:circuits}) and so stay on its manifold, but a step along the Euclidean gradient $\nabla\LL_k$ with respect to $G$ would break this constraint. The return path in \cref{fig:banner}d repairs this with two geometric operations, stated here for a unitary gate $G$.

\textbf{Projection} fixes the direction. A matrix $W$ is skew-Hermitian if $W^\dagger=-W$. Differentiating $G^\dagger G=I$ along a curve of gates through $G$ with velocity $\Delta$ gives $\Delta^\dagger G+G^\dagger\Delta=0$, so $W=G^\dagger\Delta$ is skew-Hermitian. The directions that keep the constraint to first order, which form the tangent space at $G$, are therefore exactly $\Delta=G\,W$ with $W$ skew-Hermitian \citep{edelman1998geometry, absil2008optimization}. Projection keeps the gradient's nearest tangent component in the Frobenius norm, $\proj_G(\nabla\LL_k)=G\,\skew\!\bigl(G^\dagger\nabla\LL_k\bigr)$, where $\skew(A)=\tfrac12(A-A^\dagger)$ is the skew-Hermitian part of $A$, so stepping against it is steepest descent among the feasible moves.

\textbf{Retraction} fixes the step. The manifold is curved, so a finite step along a tangent direction still leaves it, and each update maps the stepped gate back. For an increment $a$ built from the projected direction, the Cayley retraction forms $W=\skew\!\bigl(a\,G^\dagger\bigr)$ and applies $G_{+}=\bigl(I-\tfrac12W\bigr)^{-1}\bigl(I+\tfrac12W\bigr)\,G$, which is exactly unitary for every skew-Hermitian $W$ and matches $G+a$ to first order.

Riemannian Adam \citep{kingma2015adam, becigneul2019riemannian} applies this sequence gate by gate, replacing the plain gradient step $a=-\eta\,\proj_G(\nabla\LL_k)$ of step size $\eta$ with an adaptive one built from moment estimates and transporting the first moment to the new tangent space by re-projection. All reported experiments use this optimizer, whose moment equations, orthogonal and phase-gate cases, and step-size schedule are given in \cref{app:riemannian}.

\begin{table*}[t!]
\centering
\caption{Mean test PSNR (dB) on $100$ held-out images per dataset (image size in parentheses) at keep ratios $\rho$. \textit{Ours} are the trained bases, \textit{Classical} the fixed $8\times8$ block transforms. Per column, the highest mean is \textbf{bold} and the second highest \underline{underlined}. MERA is undefined on Quick Draw's $5$ qubits per axis (---). Per-image PSNRs, standard errors ($0.1$--$1.0$~dB per cell) and paired-bootstrap intervals are archived with the benchmarks \citep{an2026benchmarks}.}
\label{tab:div2k_repr}\label{tab:quickdraw_repr}
\small
\setlength{\tabcolsep}{3.6pt}
\begin{tabular}{@{}l ccccc ccccc@{}}
\toprule
& \multicolumn{5}{c}{\textbf{DIV2K} ($256^2$)} & \multicolumn{5}{c}{\textbf{Quick Draw} ($32^2$)} \\
\cmidrule(lr){2-6}\cmidrule(l){7-11}
$\rho = $ & 0.01 & 0.05 & 0.10 & 0.20 & 0.40 & 0.01 & 0.05 & 0.10 & 0.20 & 0.40 \\
\midrule
\multicolumn{11}{@{}l}{\textit{Ours}} \\
QFT            & 21.00 & 24.68 & 27.14 & 30.84 & 37.54 & 12.87 & 16.43 & 19.22 & 23.82 & 33.51 \\
Entangled QFT  & 21.00 & 24.68 & 27.14 & 30.84 & 37.54 & 12.87 & 16.43 & 19.22 & 23.82 & 33.51 \\
RichBasis      & 14.54 & 25.67 & \underline{28.84} & \underline{33.31} & \underline{40.93} & \underline{13.42} & \underline{18.57} & \underline{22.73} & \underline{29.59} & \underline{44.11} \\
TEBD           & 20.53 & 25.69 & 28.63 & 32.91 & 40.37 & \textbf{13.43} & 18.55 & 22.69 & 29.56 & 43.88 \\
MERA           & \underline{21.30} & \underline{25.77} & 28.60 & 32.78 & 40.09 & --- & --- & --- & --- & --- \\
DCT-IV         & \textbf{21.92} & \textbf{26.11} & 28.82 & 32.77 & 39.60 & 13.08 & \textbf{18.92} & \textbf{23.79} & \textbf{31.57} & \textbf{44.62} \\
\multicolumn{11}{@{}l}{\textit{Classical}} \\
DFT ($8{\times}8$) & 14.31 & 24.09 & 26.70 & 30.37 & 36.72 & 12.42 & 15.82 & 18.55 & 22.81 & 32.25 \\
DCT-II ($8{\times}8$) & 14.32 & 25.71 & \textbf{28.98} & \textbf{33.59} & \textbf{41.43} & 12.70 & 16.89 & 20.36 & 26.12 & 39.00 \\
\bottomrule
\end{tabular}
\end{table*}

\subsection{Initialization}
\label{sec:train_robustness}

Initialization is topology-specific: QFT and DCT-IV begin at their defining fixed-transform values, RichBasis at the same QFT point, Entangled QFT at the separable QFT with small seeded pair couplings, and TEBD and MERA at seeded feasible parameters. Because the objective is non-convex, these choices do not provide a global-optimality guarantee, and randomly initialized parameterized circuits are known to face flat optimization landscapes \citep{mcclean2018barren, cerezo2021cost}. The QFT gate-unfreezing random-seed tests and DCT-IV perturbation studies in \cref{app:direct_training,app:exact_disturbance} empirically describe the initializations considered.

\section{Experiments}
\label{sec:experiments}

The training framework on the full isometric tensor-network family is tested on DIV2K and Quick Draw line drawings. DIV2K images are the high-resolution split, converted to grayscale, center-cropped to a square, and Lanczos-resampled to $256 \times 256$. Quick Draw images are the dataset's $28 \times 28$ rasterized bitmaps zero-padded to $32 \times 32$. Image pixels in both DIV2K and Quick Draw are scaled to $[0,1]$. Every dataset contributes a fixed $500$-image training slice and $100$ test images, drawn once with a fixed seed shared by every method.

\begin{figure}[!tbp]
\centering
\includegraphics[width=\columnwidth]{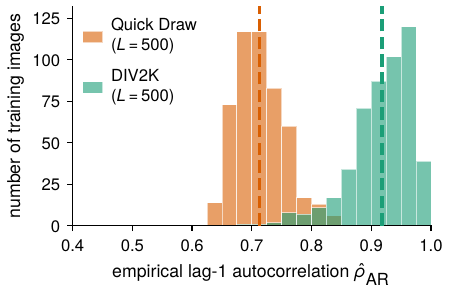}
\caption{Empirical lag-1 autocorrelation $\hat{\rho}_{\mathrm{AR}} = \tfrac{1}{2}(\hat{\rho}_{\mathrm{row}} + \hat{\rho}_{\mathrm{col}})$ across each $500$-image training slice. Dashed vertical lines mark the dataset means}
\label{fig:ar1_histogram}
\end{figure}

All learned bases are trained at $\rho = 0.1$ and evaluated at five keep ratios. The two datasets sit at opposite sides of the AR(1)--Gaussian family with covariance $R_{ij} = \sigma_{\mathrm{AR}}^2 \rho_{\mathrm{AR}}^{\,|i-j|}$. The measured lag-1 autocorrelation (\cref{fig:ar1_histogram}) places DIV2K near the $\hat{\rho}_{\mathrm{AR}} \to 1$ limit, where the DCT-II is asymptotically optimal (\cref{app:dct_suboptimality}), and Quick Draw well below it. Quality is reported as the peak signal-to-noise ratio $\mathrm{PSNR} = 10\log_{10}(1/\mathrm{MSE})$ in dB computed for each of the $100$ test images, with the mean squared error $\mathrm{MSE} = \LL_k/2^{m+n}$ the loss per pixel, and reconstructions clipped to the real pixel range before scoring. \Cref{tab:div2k_repr} benchmarks the family against JPEG's $8 \times 8$ block DCT-II and DFT references. End-to-end learned codecs are excluded because their bits-per-pixel rate is not comparable to $\rho$.

\subsection{Topology comparison}
\label{sec:topology_div2k}

\begin{figure*}[t!]
\centering
\includegraphics[width=\textwidth]{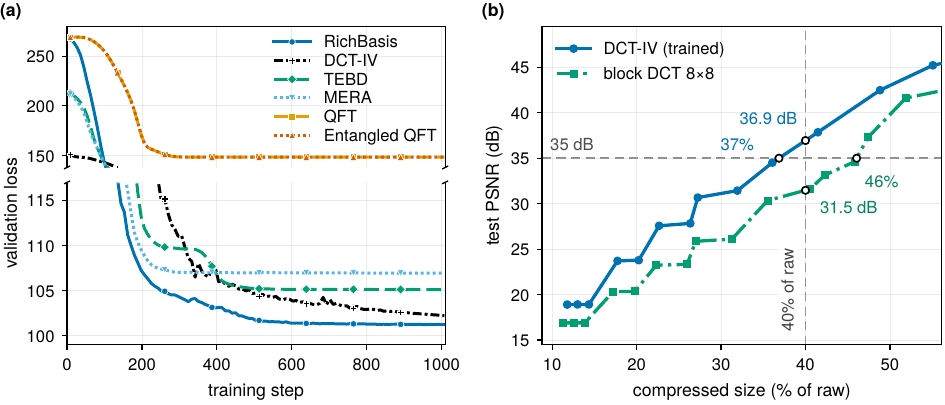}
{\phantomsubcaption\label{fig:topology_loss}}
{\phantomsubcaption\label{fig:rd_quickdraw}}
\caption{Training and compression of the learned bases. (a)~Validation loss on DIV2K (broken vertical axis). (b)~Quick Draw rate--distortion under the byte-level codec. Open circles mark crossings with the dashed guides. At $35$~dB the trained DCT-IV needs $37\%$ of raw against the $8 \times 8$ block DCT's $46\%$ ($20\%$ fewer bytes), and at $40\%$ of raw it reaches $36.9$~dB against $31.5$~dB}
\label{fig:loss_and_rd}
\end{figure*}

We compare the six learned bases against the classical references on both datasets in \cref{tab:div2k_repr}. MERA needs a power-of-two qubit count per axis, so five compete on Quick Draw.

The DCT-IV has the highest learned-basis mean at two of the five DIV2K keep ratios and four of the five Quick Draw ones, while RichBasis takes the remaining DIV2K means. One exception is sharp rather than marginal. At DIV2K $\rho = 0.01$ RichBasis falls to $14.54$~dB while DCT-IV reaches $21.92$~dB. Retaining $1\%$ of the coefficients is a tenfold extrapolation beyond the $\rho = 0.1$ budget the bases were trained at, and the objective is blind to the ordering \emph{within} the retained set. Averaged over test images, the retained energy fraction $\lVert\topk(\TT\,\mathbf{x},\,k)\rVert_F^2/\lVert\mathbf{x}\rVert_F^2 = 1-\LL_k/\lVert\mathbf{x}\rVert_F^2$ exceeds $99\%$ for both bases at their top $10\%$ of coefficients, but the DCT-IV, anchored at the cosine transform's steeply ordered spectrum, concentrates $96\%$ into its top $1\%$ where the freely trained RichBasis keeps only $84\%$.

Across the whole table, the local gate manifold separates the learned bases more than the wiring does. TEBD, MERA, and RichBasis carry the same general two-qubit $U(4)$ gate under ring, hierarchical, and all-to-all wiring, yet sit within $0.9$~dB of one another at every DIV2K ratio above $0.01$. The DCT-IV, whose controlled $X$ gates relax over the full $O(4)$ rather than a diagonal manifold in \cref{sec:alt_topologies}, lands in the same group. QFT and Entangled QFT, which keep the diagonal $U(1)^4$ phase, fall $1.0$ to $3.4$~dB below that group on DIV2K as $\rho$ grows, and up to $11.1$~dB on Quick Draw. They are numerically tied: the coupling, applied after the transform, cannot alter magnitude-top-$k$ reconstruction. \Cref{fig:topology_loss} shows the same split in the validation loss $\frac{1}{|\mathcal{V}|}\sum_{\mathbf{x}\in\mathcal{V}}\LL_k(\theta;\mathbf{x})$, the $|\mathcal{V}| = 75$ images held out from each training slice: the four dense-gate bases floor within $6$ of one another and the diagonal pair floors roughly $45$ above the group mean.

Against the classical references, every learned basis beats the $8\times8$ DFT at every keep ratio on both datasets. The cosine reference is the harder one. On Quick Draw the three dense-gate bases (RichBasis, TEBD, DCT-IV) beat the $8\times8$ DCT-II at every ratio, by up to $5.6$~dB, while QFT and Entangled QFT lead it only at $\rho = 0.01$. On AR(1)-like DIV2K the best learned basis leads only under aggressive truncation and trails from $\rho = 0.10$ upward. The learned bases pay off where energy compaction is scarcest, the regime the next section prices in bytes.

\begin{figure*}[!tbp]
\centering
\includegraphics[width=\textwidth]{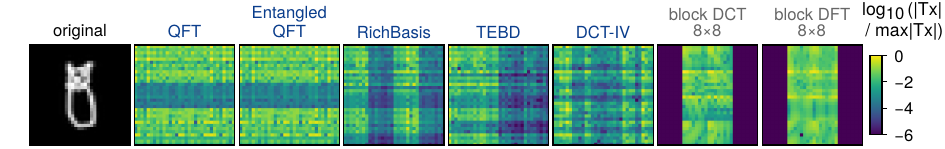}\\[2pt]
\includegraphics[width=\textwidth]{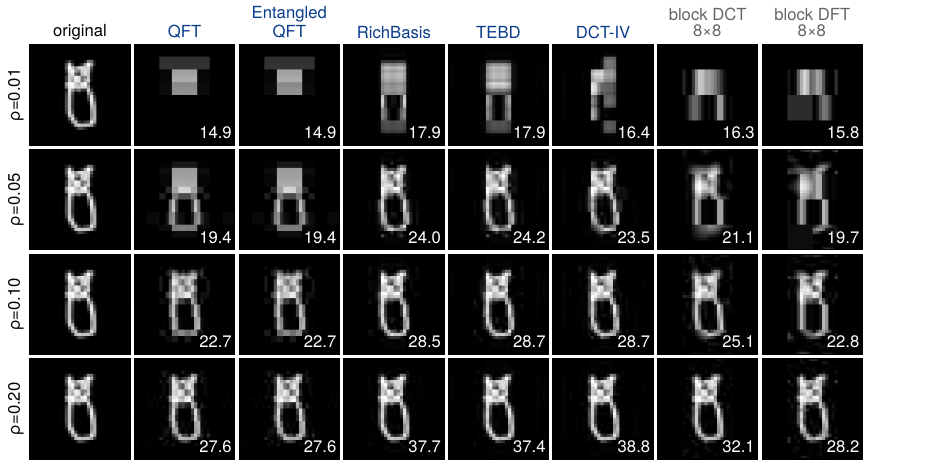}
\caption{Frequency-domain coefficient magnitude (top, $\log_{10}\bigl(\lvert\TT\mathbf{x}\rvert/\max\lvert\TT\mathbf{x}\rvert\bigr)$, peak-normalized per panel) and reconstructions at keep ratios $\rho$ (per-cell PSNR in dB) for a Quick Draw cat sketch ($32 \times 32$). The leftmost cell is the original. The columns are the five trained bases (QFT, Entangled QFT, RichBasis, TEBD, DCT-IV) in the order of \cref{tab:div2k_repr}, then the $8\times8$ classical DCT and DFT references, with learned bases in blue and classical references in gray. MERA is omitted, as Quick Draw's qubit count per axis, $5$, is not a power of two. Image from Quick Draw (\citealp{quickdraw}; CC BY 4.0)}
\label{fig:freqrecon_quickdraw}
\end{figure*}

\begin{figure*}[!tbp]
\centering
\includegraphics[width=\textwidth]{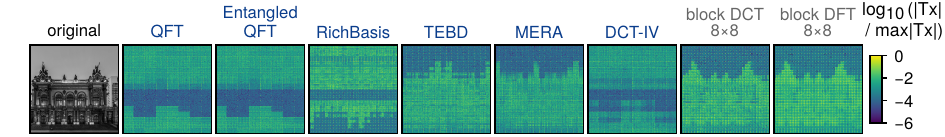}\\[2pt]
\includegraphics[width=\textwidth]{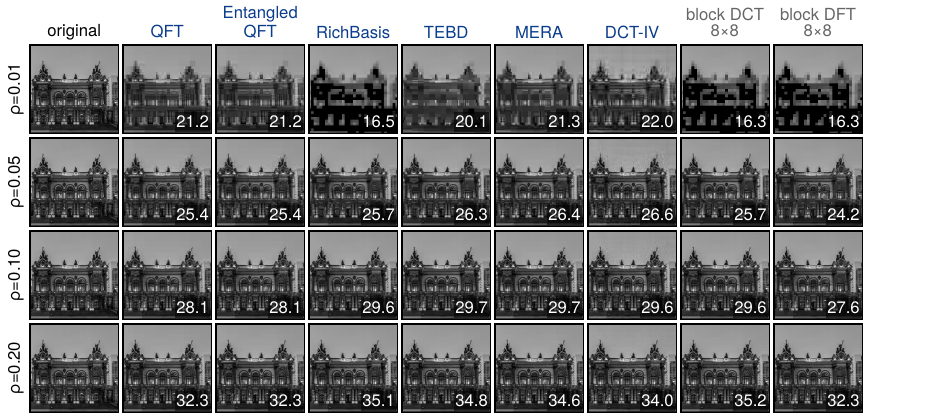}
\caption{Frequency-domain coefficient magnitude (top, $\log_{10}\bigl(\lvert\TT\mathbf{x}\rvert/\max\lvert\TT\mathbf{x}\rvert\bigr)$, peak-normalized per panel) and reconstructions at keep ratios $\rho$ (per-cell PSNR in dB) for a DIV2K fa\c{c}ade ($256 \times 256$, drawn from the training slice). The leftmost cell is the original. The columns are all six trained bases of \cref{tab:div2k_repr}, including MERA, in its order, then the $8\times8$ classical DCT and DFT references, with learned bases in blue and classical references in gray. Image from DIV2K \citep{div2k}}
\label{fig:freqrecon_div2k}
\end{figure*}

The example reconstructions in \cref{fig:freqrecon_quickdraw,fig:freqrecon_div2k} add what the table's mean PSNRs cannot show. The dense-gate bases preserve the sketch strokes that the diagonal pair blurs, and on DIV2K the learned circuits, apart from the RichBasis collapse at $\rho = 0.01$ noted above, avoid the block artifacts of the $8 \times 8$ references under aggressive truncation.

\subsection{Compression at matched quality}
\label{sec:compression}

\Cref{fig:rd_quickdraw} prices the reconstruction quality of \cref{tab:quickdraw_repr} in bytes, and the trained basis wins on both axes of the rate--distortion plane. Horizontally, at matched quality, reaching $35$~dB costs the trained DCT-IV $20\%$ fewer bytes than JPEG's fixed $8 \times 8$ block DCT (about $30\%$ fewer near $27$~dB). Vertically, at matched size, the trained basis gains $+5.4$~dB at $40\%$ of raw ($8$~bits per pixel), so at equal storage the block DCT's mean-squared pixel error is about $3.5$ times larger. These two operating points are our main compression result.

The comparison holds the codec fixed and changes only the basis. Each image is transformed, the $k$ largest coefficients are kept and uniformly quantized to $b$ bits, and the values and positions are entropy-coded into a file, the standard transform-coding chain, with the byte layout in the benchmark repository. The basis itself is one parameter file per dataset and is excluded from the reported sizes, as are the transform definitions of the classical references. Sweeping the keep ratio and the bit depth $b \in \{6, 8, 10\}$ traces one curve per plotted basis.

A possible explanation is the spatial extent of the content. A sketch stroke runs across the frame while an $8 \times 8$ block sees it only in fragments, so a full-image basis may reach a matched quality with fewer coefficients, each of which would otherwise be stored. The saving is essentially the same when the stored corpus is the training corpus itself ($19.7\%$ at $35$~dB). Throughout, the comparison is basis against basis under the same lossy codec and budget, not between complete codecs.

\section{Conclusion}
\label{sec:conclusion}

\subsection{Summary}

Our contribution is a trainable family of unitary image bases built from the classical fast transforms. Reading the FFT and the DCT-IV as isometric tensor networks and relaxing each gate within its matrix manifold gives bases that contain these transforms as exact points, apply in near-linear time, and add only polylogarithmically many parameters. We train one shared basis per dataset with Riemannian optimization, which keeps every gate unitary throughout training, so each trained basis remains invertible.

The experiments show that, within the tested family, the local gate manifold matters more than the wiring: the dense-gate bases lead the diagonal-phase pair at every keep ratio from $0.05$ upward, while rewiring the same $U(4)$ gate changes the DIV2K means by less than $1$~dB above $\rho = 0.01$. Against the $8\times8$ block DCT, the outcome depends on the dataset. On Quick Draw, DCT-IV, TEBD, and RichBasis have a higher mean PSNR at all five tested keep ratios, while QFT and Entangled QFT lead only at $\rho = 0.01$. On DIV2K, the best learned mean leads only at the two lowest keep ratios. Under a common codec, the trained DCT-IV stores Quick Draw images in about $20\%$ fewer bytes at $35$~dB, excluding the one-off storage of its parameters.

\subsection{Outlook and limitations}
\label{sec:discussion}

The limited effect of wiring changes in the tested family (\cref{sec:topology_div2k}) motivates exploring other training objectives, and the trained bases may also serve tasks beyond compression.

On the optimization side, the objective can move beyond hard top-$k$ truncation. The loss of \cref{eq:sparse_basis_problem} is blind to the ordering within the retained set (the RichBasis drop at $\rho = 0.01$ in \cref{sec:topology_div2k}) and never sees the quantizer or the entropy coder that the codec of \cref{sec:compression} applies afterward. The hard selection could give way to soft thresholding, the proximal operator of the $\ell_1$ norm \citep{donoho1995denoising}, which shrinks every coefficient toward zero by a threshold instead of keeping exactly $k$. The reconstruction then varies continuously with the coefficients, and training targets an $\ell_1$-type sparsity rather than a fixed budget. The squared error has many alternatives as well. Objectives averaged over several keep ratios, perceptual criteria such as the structural similarity index \citep{wang2004ssim}, and rate--distortion losses evaluated on the coded byte stream would each train the basis against a quantity closer to how it is used. Training cost has similar room. Each step pays the gate-by-gate contraction cost of \cref{app:transform_complexity}, and fusing recursion levels where the two-leg gates stay diagonal, or optimizing the contraction order outright \citep{liu2021tropical, liu2023generic}, would lower it toward the fast-transform schedule and admit larger registers and deeper topologies.

Beyond compression, Fourier-type bases serve several other tasks that a trained basis could take on. A related trainable relaxation of the QFT already outperforms fixed transforms in image inpainting \citep{an2026inpainting}. The coefficients of a trained basis also give a data-adapted frequency description of a dataset for spectral analysis, as in \cref{fig:freqrecon_quickdraw,fig:freqrecon_div2k}. Fourier neural operators apply a fixed FFT in each layer \citep{li2021fourier}, which a trained fast basis could replace. Finally, each basis is itself a quantum circuit on $m+n$ qubits, so it could act as a trained circuit in quantum kernel methods \citep{havlicek2019supervised, schuld2019quantum}. Since a unitary alone leaves the overlap of two states unchanged, such a kernel would need to compare the states after measurement or truncation.

Two limitations bound the result. Each basis is trained offline for one dataset, so the method is a basis-design tool rather than a deployable codec, and the topology is fixed in advance. Comparisons against wavelets, including circuit-designed wavelet families \citep{evenbly2018wavelets, mccord2022wavelets}, and against learned dictionaries would place the family within the broader transform-coding toolbox. Other corpora that sit away from the AR(1) regime, such as diagrams or the frames of drawn and animated video, are natural targets for the same train-once recipe but remain untested.

\FloatBarrier
\backmatter

\bmhead{Acknowledgements}
We thank Lei Wang for the helpful discussion about the JPEG algorithm. The authors are grateful to the Institute of Science Tokyo for providing the research environment.

\section*{Statements and Declarations}

\bmhead{Funding}
This work was supported in part by the National Key R\&D Program of China (Grant No.~2024YFB4504004), the National Natural Science Foundation of China (Grant No.~12404568), and the Japan Science and Technology Agency (JST) SPRING program (Grant No.~JPMJSP2180).

\bmhead{Competing interests}
The authors have no relevant financial or non-financial interests to disclose.

\bmhead{Ethics approval and consent to participate}
Not applicable.

\bmhead{Consent for publication}
Not applicable.

\bmhead{Data availability}
The DIV2K \citep{div2k} and Quick Draw \citep{quickdraw} datasets used in this study are publicly available. The trained bases and per-image evaluation records are archived on Zenodo as version 0.1.0 of the benchmark repository \citep{an2026benchmarks}, which is developed at \url{https://github.com/zazabap/pdft-benchmarks}.

\bmhead{Materials availability}
Not applicable.

\bmhead{Code availability}
The open-source implementation \texttt{pdft} is archived on Zenodo as version 0.3.0 \citep{an2026pdft}, the version behind every reported result, and is developed at \url{https://github.com/zazabap/pdft}. The benchmark suite with its reproduction scripts is archived with the data \citep{an2026benchmarks}.

\bmhead{Author contributions}
Shiwen An, as first author, was involved in every part of the project, including the simulations, the code, and drafting the manuscript. Zhongyi Ni completed the prototype and reviewed the theory. Huanhai Zhou assisted with the code and the benchmark validation. Jin-Guo Liu supervised the project, guiding it with instructions and revisions throughout. All authors read and approved the final manuscript.

\bmhead{Use of AI tools}
AI-assisted tools (Claude, Anthropic; and Codex, OpenAI) were used for language editing and for consistency checks across the manuscript, references, code, and archived outputs. All source verification, statistical and scientific decisions, and final content were independently reviewed and approved by the authors, who take full responsibility for the work.

\begin{appendices}

\section{Network derivations}
\label{app:decompose}

\begin{figure*}[!tbp]
  \centering
  \includegraphics[width=\textwidth]{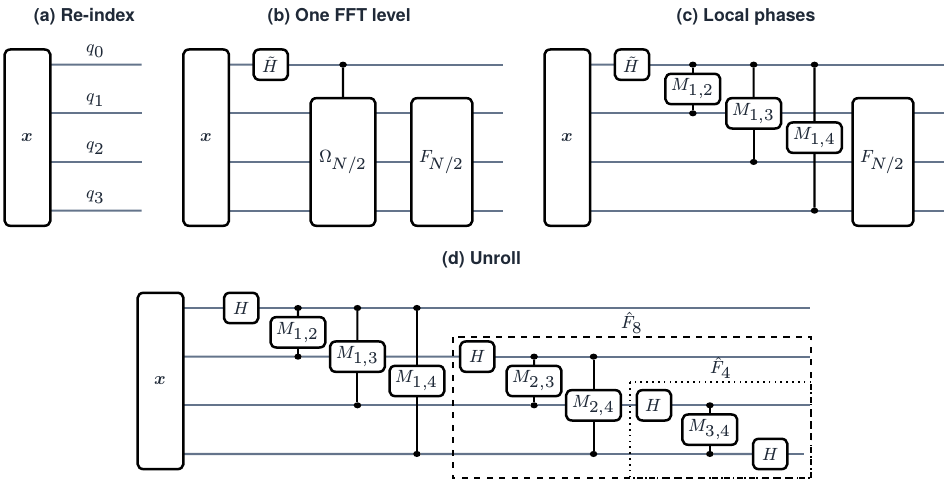}
  \caption{FFT-to-QFT construction for $N=16$: (a) binary re-indexing, (b) one Cooley--Tukey level, (c) local phase factors, and (d) recursive unrolling and relaxation. Panels (a)--(c) preserve the matrix. Panel (d) normalizes by $N^{-1/2}$ and relaxes each gate within its manifold. Here $\tilde H$ is the unnormalized Hadamard, $\Omega_{N/2}$ the twiddle factors, and $M_{i,j}$ the controlled phases. Dashed and dotted boxes mark the normalized children $\widehat F_8$ and $\widehat F_4$ ($\widehat F_N=N^{-1/2}F_N$), and fixed output reordering is omitted}
  \label{fig:cooley_tukey_to_qft}
\end{figure*}

This appendix derives the parametric tensor networks trained in the main text.
Starting from the Cooley--Tukey FFT, we apply local transformations so that every remaining local gate is a free element of a unitary or orthogonal manifold. The \cref{fig:cooley_tukey_to_qft} shows the four transformations for an $n=4$-leg example.

\subsection{FFT decomposition}
\label{app:gate_decomposition}

The DFT matrix of size $N = 2^{n}$ has entries $(F_N)_{i,j} = \omega^{(i-1)(j-1)}$ with $\omega = e^{-2\pi i / N}$ the primitive $N$-th root of unity, so $F_N$ is the Vandermonde matrix
\begin{equation}
  F_N = \begin{pmatrix}
    1 & 1 & 1 & \cdots & 1 \\
    1 & \omega & \omega^2 & \cdots & \omega^{N-1} \\
    1 & \omega^2 & \omega^4 & \cdots & \omega^{2(N-1)} \\
    \vdots & \vdots & \vdots & \ddots & \vdots \\
    1 & \omega^{N-1} & \omega^{2(N-1)} & \cdots & \omega^{(N-1)^2}
  \end{pmatrix}.
\end{equation}
Split the columns (the input index) at $N/2$ into a top half $\mathbf{x}_{\text{top}}$ and a bottom half $\mathbf{x}_{\text{bot}}$, and the rows (the output index) by the parity of the output frequency $i-1$. Since $\omega^{(i-1)(j + N/2 - 1)} = (-1)^{i-1}\,\omega^{(i-1)(j-1)}$, the rows computing even output frequencies add the two column halves and those computing odd frequencies subtract them, and each of the four resulting $(N/2) \times (N/2)$ blocks reduces to the half-size DFT up to the twiddle $\Omega_{N/2} = \diag(1, \omega, \ldots, \omega^{N/2-1})$:
\begin{equation}
  \begin{aligned}
    {[F_N]}^{\text{even}}_{\text{left}} &= {[F_N]}^{\text{even}}_{\text{right}} = F_{N/2}, \\
    {[F_N]}^{\text{odd}}_{\text{left}} &= -{[F_N]}^{\text{odd}}_{\text{right}} = F_{N/2}\, \Omega_{N/2},
  \end{aligned}
  \label{eq:blocks}
\end{equation}
using $(\omega^2)^{(i-1)(j-1)} = (F_{N/2})_{i,j}$ on both row parities, $\omega^N = 1$ on the right-even block, and $\omega^{N/2} = -1$ on the right-odd block.
Assembling the four blocks gives the Cooley--Tukey factorization
\begin{equation}
  F_N \mathbf{x}
  = P_N \!
    \begin{pmatrix} F_{N/2} & 0 \\ 0 & F_{N/2} \end{pmatrix}
    \!\!
    \begin{pmatrix} I_{N/2} & I_{N/2} \\ \Omega_{N/2} & -\Omega_{N/2} \end{pmatrix}
    \!\!
    \begin{pmatrix} \mathbf{x}_{\text{top}} \\ \mathbf{x}_{\text{bot}} \end{pmatrix}.
  \label{eq:fft}
\end{equation}
Here $P_N$ is the one-level perfect-shuffle interleaver. If
$\mathbf{z}=(\mathbf{z}_{\mathrm e}^{\mathsf T},\mathbf{z}_{\mathrm o}^{\mathsf T})^{\mathsf T}$ stores all even outputs before all odd outputs, then
$(P_N\mathbf{z})_{2r}=(z_{\mathrm e})_r$ and $(P_N\mathbf{z})_{2r+1}=(z_{\mathrm o})_r$ for $0\leq r<N/2$.
Thus $P_N$ only restores natural output order at this level; the interleavers accumulated over all recursion levels compose into the full bit reversal of the standard QFT.
Applied recursively, each level reduces an $N$-point DFT to two $(N/2)$-point DFTs plus $O(N)$ twiddle multiplications, giving total cost $O(N \log N)$.
\Cref{eq:fft} is the input to the tensor-network construction.

\paragraph{Matrix-to-network construction.}
The remaining steps preserve the matrix in \cref{eq:fft} while exposing its factors as local gates in \cref{fig:cooley_tukey_to_qft}.

\textbf{Re-index} (\cref{fig:cooley_tukey_to_qft}a). A vector of length $N=2^n$ carries a single integer index that we write in most-significant-bit-first order as $i=2^{n-1}q_0+2^{n-2}q_1+\cdots+q_{n-1}$.
Re-interpreting $\mathbf{x}$ as a rank-$n$ tensor with one binary leg $q_j$ per bit is purely notational, but it turns every linear operator in \cref{eq:fft} into a tensor-network component acting on specified legs.

\textbf{Recurse} (\cref{fig:cooley_tukey_to_qft}b). Take $q_0$ as the most-significant bit splitting \cref{eq:fft} into its top and bottom halves. The combined butterfly--twiddle on the input side factors through the identity
\begin{equation}
  \begin{pmatrix} I_{N/2} & I_{N/2} \\ \Omega_{N/2} & -\Omega_{N/2} \end{pmatrix}
  = \begin{pmatrix} I_{N/2} & 0 \\ 0 & \Omega_{N/2} \end{pmatrix}
    (\tilde H \otimes I_{N/2}),
  \label{eq:butterfly-id}
\end{equation}
with $\tilde H = \bigl(\begin{smallmatrix} 1 & 1 \\ 1 & -1 \end{smallmatrix}\bigr)$ the unnormalized Hadamard matrix. Substituting \cref{eq:butterfly-id} into \cref{eq:fft} and reading $q_0$ as the leading binary leg gives the single-level factorization described in words in \cref{sec:qft},
\begin{equation}
  F_N = P_N \, (I_2 \otimes F_{N/2}) \begin{pmatrix} I_{N/2} & 0 \\ 0 & \Omega_{N/2} \end{pmatrix} (\tilde H \otimes I_{N/2}),
  \label{eq:ck_factorization}
\end{equation}
reading \cref{eq:ck_factorization} right to left: a Hadamard factor on $q_0$, a diagonal twiddle controlled by that leg, the half-size DFT $F_{N/2}$ on $(q_1,\ldots,q_{n-1})$, and the one-level interleaver $P_N$ returning the outputs to natural order.

\textbf{Decompose} (\cref{fig:cooley_tukey_to_qft}c). The controlled twiddle factors into a chain of two-leg phase gates,
\begin{equation}
  \begin{aligned}
  \begin{pmatrix} I_{N/2} & 0 \\ 0 & \Omega_{N/2} \end{pmatrix}
    &= \prod_{j=1}^{n-1} \mathrm{CP}_{0,j}(\Lambda_j),\\
  \Omega_{N/2}
    &=\diag(1,\omega^{N/4})\otimes\cdots\otimes\diag(1,\omega).
  \end{aligned}
\end{equation}
Thus the twiddle is a chain of $n-1$ diagonal gates $\mathrm{CP}_{0,j}(\Lambda_j)$, each joining $q_0$ to a lower-significance leg $q_j$ with the $2\times2$ block $\Lambda_j=\diag\bigl(1,\omega^{2^{\,n-1-j}}\bigr)$. These are the level-one instances of the designed controlled phases of \cref{sec:qft}: since leg $q_j$ is bit $j+1$, the gate $\mathrm{CP}_{0,j}(\Lambda_j)$ is, up to the conjugation fixed below, the controlled phase $M_{1,j+1}$ drawn in \cref{fig:banner}b, and the recursion on $F_{N/2}$ supplies the pairs $M_{i,j}$ with $i>1$. Here $\omega=e^{-2\pi i/N}$, while the quantum-circuit convention of \cref{sec:qft} and the \texttt{pdft} implementation use the conjugate root $e^{2\pi i/N}$, under which the identical derivation yields the positive phases $\phi_{i,j}=2\pi/2^{\,j-i+1}$.

\textbf{Relax} (\cref{fig:cooley_tukey_to_qft}d). Recursing on $F_{N/2}$ replays the butterfly-and-decomposition pair on $(q_1,\ldots,q_{n-1})$, down to $F_2=\tilde H$ on the final leg. The fixed per-level interleavers compose across the recursion into the usual output bit reversal. For the unitary QFT, normalize $F_N$ to $\widehat F_N=N^{-1/2}F_N$, so each butterfly is the normalized Hadamard $H=2^{-1/2}\bigl(\begin{smallmatrix}1&1\\1&-1\end{smallmatrix}\bigr)$ of \cref{sec:qft}. Relaxation then replaces each fixed local gate by a trainable element of its manifold (\cref{tab:gate_relaxations}). The resulting network $\TT(\theta)$ contains the normalized DFT at the Fourier point and is unitary at every parameter setting.

\subsection{DCT-IV decomposition}
\label{app:dct4_circuit}

\begingroup
\setlength{\abovedisplayskip}{5pt plus 2pt minus 2pt}
\setlength{\belowdisplayskip}{5pt plus 2pt minus 2pt}
\setlength{\abovedisplayshortskip}{3pt plus 1pt minus 1pt}
\setlength{\belowdisplayshortskip}{4pt plus 1pt minus 1pt}
\renewcommand{\arraystretch}{0.92}

For $N=2^n$, the orthonormal DCT-IV has entries
$(C_N^{\mathrm{IV}})_{u,v}=\sqrt{2/N}\cos\psi_{u,v}$, with kernel angle
$\psi_{u,v}=\pi(2u+1)(2v+1)/(4N)$.
We expose its radix-2 factorization directly as a tensor network on binary
wires $q_0,q_1,\ldots,q_{n-1}$; $q_0$ selects the recursion branch and the
remaining wires carry the child index. Radix-2 DCT-IV factorizations are classical; we use the recursive sparse form from the factorization survey \citep{plonka2005dct}, read within the algebraic Cooley--Tukey framework \citep{puschel2008asp}.

\begin{figure*}[!tbp]
  \centering
  \includegraphics[width=\textwidth]{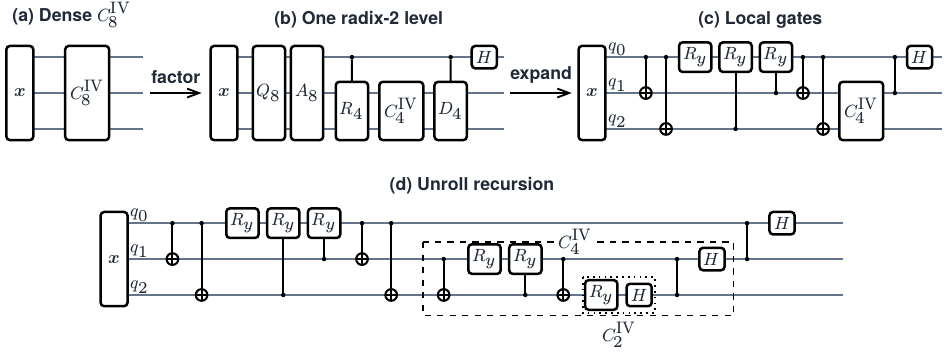}
  \caption{DCT-IV tensor-network decomposition at $N=8$. (a)~The starting dense matrix $C_8^{\mathrm{IV}}$. (b)~One radix-2 level, equal to $C_8^{\mathrm{IV}}$ up to output order: $Q_8$ pairs each input with its mirror, $A_8$ rotates each pair, $R_4$ reverses the lower branch, both branches run the same child $C_4^{\mathrm{IV}}$, $D_4$ applies the sign correction, and $H$ merges the branches. (c)~The local forms of $Q_8$, $A_8$, $R_4$, and $D_4$ as one- and two-qubit gates. (d)~The recursion unrolled into an exact $R_y$ network. Dashed and dotted boxes mark the children $C_4^{\mathrm{IV}}$ and $C_2^{\mathrm{IV}}$, and the trainable manifolds are listed in \cref{tab:gate_relaxations}. Dots mark controls, $\oplus$ marks a CNOT target, and a dot pair is a controlled $Z$. Fixed output wiring is omitted}
  \label{fig:cooley_tukey_to_dct}
\end{figure*}

\paragraph{Two kernel identities.}
The construction rests on two elementary facts about the kernel angle
$\psi_{u,v}$, both following directly from its definition. First, an input
and its mirror collapse together,
\begin{equation}
  \cos\psi_{u,\,N-1-v}=(-1)^u\sin\psi_{u,v},
  \label{eq:dct4_mirror}
\end{equation}
so $x_v$ and $x_{N-1-v}$ reach each output $u$ through a single sine--cosine
pair whose sine term carries the $(-1)^u$ sign; this is why $Q_N$ below pairs $v$ with $N-1-v$ and
why one branch sign accounts for the whole mirror. Second, the even and odd
parent outputs $2w$ and $2w+1$ that share a child output $w$ differ only by a
local offset,
\begin{equation}
  \psi_{2w,v}=\alpha_{w,v}-\delta_v,\qquad
  \psi_{2w+1,v}=\alpha_{w,v}+\delta_v,
  \label{eq:dct4_split}
\end{equation}
with $\alpha_{w,v}=\tfrac{\pi(2w+1)(2v+1)}{2N}$ the length-$N/2$ DCT-IV kernel
angle and $\delta_v=\tfrac{\pi(2v+1)}{4N}$ independent of $w$. A single rotation
by $\delta_v$ on the input pair therefore prepares both outputs, and the shared
child $C_{N/2}^{\mathrm{IV}}$ supplies the common $\alpha_{w,v}$ projection.

\paragraph{Matrix-to-network construction.}
One radix-2 level factors the transform as
\begin{equation}
  \begin{aligned}
  C_N^{\mathrm{IV}}
  &=P_N(H\otimes I_{N/2})(I_{N/2}\oplus D_{N/2})
    (I_2\otimes C_{N/2}^{\mathrm{IV}})\\[-1pt]
  &\qquad(I_{N/2}\oplus R_{N/2})\,A_NQ_N.
  \end{aligned}
  \label{eq:dct4_recursion}
\end{equation}
Read right to left, the factors pair and rotate the inputs, run the same
half-size child on both branches, and merge the branches back into natural
output order (\cref{fig:cooley_tukey_to_dct}b). The stages below define each
factor and reduce it to one- and two-qubit gates while preserving the DCT-IV
matrix.

\textbf{Re-index} (\cref{fig:cooley_tukey_to_dct}a--b). Intermediate vectors use branch-major
order $(q_0,v)$: positions $v$ and $v+N/2$ carry the two recursion branches for
the same child index $v$. For $0\leq v<N/2$, define
\begin{equation*}
  (R_{N/2}\mathbf z)_v=z_{N/2-1-v},\qquad
  Q_N=I_{N/2}\oplus R_{N/2},
\end{equation*}
\begin{equation*}
  (P_N\mathbf z)_{2v}=z_v,\qquad
  (P_N\mathbf z)_{2v+1}=z_{v+N/2}.
\end{equation*}
Thus $Q_N$ pairs each input with its mirror, and $P_N$ interleaves the two
output branches. Both are fixed permutations. The recursion's separate factor
$I_{N/2}\oplus R_{N/2}$ is the same matrix as $Q_N$, kept written out because
it plays the branch-reversal role there rather than the pairing role.

\textbf{Rotate} (\cref{fig:cooley_tukey_to_dct}b). This factor applies the
offset $\delta_v$ of the parity split \cref{eq:dct4_split} to each pair. With
$\Gamma_{N/2}=\diag(\cos\delta_v)$ and $\Sigma_{N/2}=\diag(\sin\delta_v)$, the
block-diagonal pair rotation is
\begin{equation*}
  A_N=
  \begin{pmatrix}
    \Gamma_{N/2}&-\Sigma_{N/2}\\
    \Sigma_{N/2}&\Gamma_{N/2}
  \end{pmatrix},
\end{equation*}
with pairwise action
\begin{equation*}
  \begin{pmatrix}
    (A_N\mathbf z)_v\\ (A_N\mathbf z)_{v+N/2}
  \end{pmatrix}
  =R_y(2\delta_v)
  \begin{pmatrix}z_v\\z_{v+N/2}\end{pmatrix}.
\end{equation*}
Here $R_y(\vartheta)=\bigl(\begin{smallmatrix}
\cos(\vartheta/2)&-\sin(\vartheta/2)\\
\sin(\vartheta/2)&\cos(\vartheta/2)
\end{smallmatrix}\bigr)$. Since $\Gamma_{N/2}$ and $\Sigma_{N/2}$ are diagonal,
different child indices never mix.

\textbf{Recurse} (\cref{fig:cooley_tukey_to_dct}b). Both branches use the same
$C_{N/2}^{\mathrm{IV}}$ child, applied uncontrolled as
$I_2\otimes C_{N/2}^{\mathrm{IV}}$. The lower branch is reversed before the
child by $I_{N/2}\oplus R_{N/2}$ and receives the sign correction
$D_{N/2}=\diag((-1)^v)$ afterward. Only the shared child recurses, so no
second transform template or transform-type ancilla is required.

\textbf{Decompose} (\cref{fig:cooley_tukey_to_dct}c). The lower wires encode
$v=\sum_{p=1}^{n-1}2^{n-1-p}q_p$, so reversal complements every lower wire:
\begin{equation*}
  I_{N/2}\oplus R_{N/2}
  =\prod_{p=1}^{n-1}(I_2\oplus X)_{q_0,q_p},
\end{equation*}
a chain of controlled-NOT (CNOT) gates controlled by $q_0$, with
$X=\bigl(\begin{smallmatrix}0&1\\1&0\end{smallmatrix}\bigr)$ acting on each
lower wire as listed in \cref{tab:gate_relaxations}. Likewise, the affine angle
$2\delta_v=\pi/(2N)+\sum_{p=1}^{n-1}q_p2^{n-1-p}\pi/N$ gives
\begin{equation*}
  A_N
  =\left[\prod_{p=1}^{n-1}
    \mathrm{C}R_y\!\left(\frac{2^{n-1-p}\pi}{N}\right)_{q_p\to q_0}\right]
    R_y\!\left(\frac{\pi}{2N}\right)_{q_0}.
\end{equation*}
For $N\geq4$, the sign correction in natural child order is
\begin{equation*}
  I_{N/2}\oplus D_{N/2}=(I_2\oplus Z)_{q_0,q_{n-1}}.
\end{equation*}
Because the child's output bit reversal is deferred to the end, the parity
bit of the child output sits on the wire adjacent to $q_0$, so this
controlled $Z$ acts on $(q_0,q_1)$ as drawn in
\cref{fig:cooley_tukey_to_dct}c.

\emph{Unroll} (\cref{fig:cooley_tukey_to_dct}d). Recursively replace the shared
$C_{N/2}^{\mathrm{IV}}$ box by the same cell down to
$C_1^{\mathrm{IV}}=[1]$. The fixed $P_N$ factors accumulate into one output
bit reversal; all other gates are listed with their trainable manifolds in
\cref{tab:gate_relaxations}.

\endgroup

\subsection{Gate counts and manifolds}
\label{app:transform_complexity}
\label{app:gate_dictionary}

For a one-dimensional transform of length $N=2^n$, the unrolled QFT contains $n$ Hadamard-role gates and $n(n-1)/2$ controlled phases. Under the gate-wise relaxation of \cref{tab:gate_relaxations}, its trainable manifold therefore has $4n+4n(n-1)/2=O(n^2)=O(\log^2N)$ real parameters. The DCT-IV has the same recursion depth but a larger fixed local dictionary. At a level with $r$ lower binary legs, its two reversal fans contribute $2r$ two-leg $X$ gates, $A_N$ contributes one base $R_y$ and $r$ controlled $R_y$ gates, and the level ends with one normalized merge $H$ and, for $r>0$, one controlled $Z$. Summing over $r=0,\ldots,n-1$ gives
\begin{equation}
  \begin{aligned}
  &N_{\mathsf X}=2\sum\nolimits_{r=0}^{n-1}r=n(n-1),\qquad
   N_{\mathrm{C}R_y}=\tfrac{n(n-1)}{2},\\
  &N_{R_y}=n,\qquad N_H=n,\qquad N_{\mathsf Z}=n-1.
  \end{aligned}
  \label{eq:dct4_gate_counts}
\end{equation}

Thus both exact networks contain $O(n^2)=O(\log^2N)$ fixed-size gates. For the DCT-IV, the omitted interleavers compose into one fixed output bit reversal, and no auxiliary leg is introduced. Induction on \cref{eq:dct4_recursion}, with base case $C_1^{\mathrm{IV}}=[1]$, proves that the unrolled gates implement the orthonormal DCT-IV up to that output order.

\begin{table*}[!t]
\centering
\caption{Gate-wise relaxations of all six bases in \texttt{pdft}, with the group each gate trains on. References are exact for QFT, Entangled QFT, RichBasis, and DCT-IV, with the QFT's controlled phase between bits $i<j$ set to $\phi_{i,j}=2\pi/2^{\,j-i+1}$. TEBD and MERA list their canonical gates (the Trotter gate $e^{-i\,\delta t\,h}$ of a bond Hamiltonian $h$ over a time step $\delta t$, and the disentangler) and train from seeded values. Phase angles $\theta_r$ lie in $\RR/2\pi\mathbb{Z}$, rotation angles $\vartheta$ in $\RR/4\pi\mathbb{Z}$ with $c_\vartheta=\cos\tfrac{\vartheta}{2}$ and $s_\vartheta=\sin\tfrac{\vartheta}{2}$, and the reference angle $\vartheta_0$ is fixed at its derived value.}
\label{tab:gate_relaxations}
\tablebodyfont
\renewcommand{\arraystretch}{1.12}
\setlength{\tabcolsep}{4.5pt}
\begin{tabular*}{\textwidth}{@{\extracolsep{\fill}}lcccc@{}}
\toprule
\textbf{Gate role} & \textbf{Reference gate} & \textbf{Relaxed gate} & \textbf{Group} & \textbf{Diagram}\\
\midrule
\multicolumn{5}{@{}l}{(1)~\emph{QFT} and \emph{Entangled QFT} (which adds one matched row--column controlled phase per pair)}\\
\addlinespace[5pt]
Hadamard
& $H=\tfrac{1}{\sqrt2}\!\left(\begin{smallmatrix}1&1\\1&-1\end{smallmatrix}\right)$
& $U=\left(\begin{smallmatrix}u_{00}&u_{01}\\u_{10}&u_{11}\end{smallmatrix}\right)$
& $U(2)$
& \iconU{}\\
\addlinespace[7pt]
Controlled phase
& $M(\phi)=\left(\begin{smallmatrix}1&0&0&0\\0&1&0&0\\0&0&1&0\\0&0&0&e^{i\phi}\end{smallmatrix}\right)$
& $D(\boldsymbol\theta)=\left(\begin{smallmatrix}e^{i\theta_0}&0&0&0\\0&e^{i\theta_1}&0&0\\0&0&e^{i\theta_2}&0\\0&0&0&e^{i\theta_3}\end{smallmatrix}\right)$
& $U(1)^4$
& \iconM{}\\
\addlinespace[5pt]
\midrule
\multicolumn{5}{@{}l}{(2)~\emph{TEBD\,/\,MERA\,/\,RichBasis} ($H$ as above; ring, hierarchical, and all-to-all wiring respectively)}\\
\addlinespace[5pt]
Two-qubit gate
& \begin{tabular}{@{}c@{}}$e^{-i\,\delta t\,h}$ (TEBD)\\ disentangler (MERA)\\ $M(\phi)$ (RichBasis)\end{tabular}
& $U^{(4)}=\left(\begin{smallmatrix}u_{00}&u_{01}&u_{02}&u_{03}\\u_{10}&u_{11}&u_{12}&u_{13}\\u_{20}&u_{21}&u_{22}&u_{23}\\u_{30}&u_{31}&u_{32}&u_{33}\end{smallmatrix}\right)$
& $U(4)$
& \iconUfour{}\\
\addlinespace[5pt]
\midrule
\multicolumn{5}{@{}l}{(3)~\emph{DCT-IV}}\\
\addlinespace[5pt]
Rotation
& $R_y(\vartheta_0)=\left(\begin{smallmatrix}c_{\vartheta_0}&-s_{\vartheta_0}\\s_{\vartheta_0}&c_{\vartheta_0}\end{smallmatrix}\right)$
& $R_y(\vartheta)=\left(\begin{smallmatrix}c_\vartheta&-s_\vartheta\\s_\vartheta&c_\vartheta\end{smallmatrix}\right)$
& $O(2)$
& \iconRy{}\\
\addlinespace[7pt]
Merge
& $H=\tfrac{1}{\sqrt2}\!\left(\begin{smallmatrix}1&1\\1&-1\end{smallmatrix}\right)$
& $R_y(\vartheta)H=\tfrac{1}{\sqrt2}\!\left(\begin{smallmatrix}c_\vartheta-s_\vartheta&c_\vartheta+s_\vartheta\\c_\vartheta+s_\vartheta&s_\vartheta-c_\vartheta\end{smallmatrix}\right)$
& $O(2)$
& \iconRyH{}\\
\addlinespace[7pt]
Controlled rotation
& $I_2\oplus R_y(\vartheta_0)=\left(\begin{smallmatrix}1&0&0&0\\0&1&0&0\\0&0&c_{\vartheta_0}&-s_{\vartheta_0}\\0&0&s_{\vartheta_0}&c_{\vartheta_0}\end{smallmatrix}\right)$
& $I_2\oplus R_y(\vartheta)=\left(\begin{smallmatrix}1&0&0&0\\0&1&0&0\\0&0&c_\vartheta&-s_\vartheta\\0&0&s_\vartheta&c_\vartheta\end{smallmatrix}\right)$
& $I_2\oplus O(2)$
& \iconCRy{}\\
\addlinespace[7pt]
Controlled $Z$ (CZ)
& $I_2\oplus Z=\left(\begin{smallmatrix}1&0&0&0\\0&1&0&0\\0&0&1&0\\0&0&0&-1\end{smallmatrix}\right)$
& $D(\boldsymbol\theta)$ as above, $D(0,0,0,\pi)=I_2\oplus Z$
& $U(1)^4$
& \iconZ{}\\
\addlinespace[7pt]
Controlled $X$ (CNOT)
& $I_2\oplus X=\left(\begin{smallmatrix}1&0&0&0\\0&1&0&0\\0&0&0&1\\0&0&1&0\end{smallmatrix}\right)$
& $V=\left(\begin{smallmatrix}v_{00}&v_{01}&v_{02}&v_{03}\\v_{10}&v_{11}&v_{12}&v_{13}\\v_{20}&v_{21}&v_{22}&v_{23}\\v_{30}&v_{31}&v_{32}&v_{33}\end{smallmatrix}\right)$
& $O(4)$
& \iconCX{}\\
\addlinespace[3pt]
\bottomrule
\end{tabular*}
\end{table*}

For a $2^m\times2^n$ image the separable transforms $F_{2^m}\otimes F_{2^n}$ and $C_{2^m}^{\mathrm{IV}}\otimes C_{2^n}^{\mathrm{IV}}$ add these counts across the two registers, so both stay $O(m^2+n^2)$ in gate and parameter count. Applying the gates one at a time costs $O\!\left(2^{m+n}(m^2+n^2)\right)$ operations, the schedule the batched evaluation of \cref{app:einsum} executes; fusing each recursion level would recover the $O\!\left(2^{m+n}(m+n)\right)$ fast-transform schedule. Fusing survives relaxation only where every gate of a level stays structured, as the diagonal $U(1)^4$ bonds do; the dense $U(4)$ and $O(4)$ bonds of RichBasis and the trained DCT-IV break it, so both retain the stepped cost, as the Cost row of \cref{tab:circuits} records.

\paragraph{Relaxation manifolds.}
Every gate factor is one of three matrix manifolds, listed here with the
dimensions the parameter counts of \cref{tab:circuits} add up to:
\begin{align*}
  U(d) &= \{U\in\CC^{d\times d}:U^\dagger U=I_d\}, \\
  O(d) &= \{Q\in\RR^{d\times d}:Q^\top Q=I_d\}, \\
  U(1)^q &= \{(z_1,\ldots,z_q)\in\CC^q:|z_r|=1\},
\end{align*}
of real dimension $d^2$, $\tfrac{1}{2}d(d-1)$, and $q$ respectively, so the
factors that appear are $\dim U(2)=4$, $\dim U(4)=16$, $\dim O(2)=1$,
$\dim O(4)=6$, and $\dim U(1)^4=4$. A point of $U(1)^q$ enters the network as
the diagonal gate $\diag(z_1,\ldots,z_q)$.

The orthogonal factors are disconnected, $O(2)$ and $O(4)$ each splitting into
a $\det=+1$ and a $\det=-1$ component. The DCT-IV rotation gates and
controlled-rotation blocks start in the rotation component of $O(2)$, its merge
gates in the reflection component at $R_y(0)H=H$, and the relaxed mirror
gates in the $\det=-1$ component of $O(4)$ at the CNOT $I_2\oplus X$; because
training moves continuously, each gate stays in the component chosen at
initialization, and paired components have the same dimension.

\Cref{tab:gate_relaxations} is the complete dictionary: each fixed gate is
replaced by a trainable element of one of these manifolds with the same wiring
and arity, and the Diagram column gives the symbol that element carries in
\cref{fig:topology_circuits}. The exact Fourier phase $M(\phi)$ is the point
$\boldsymbol\theta=(0,0,0,\phi)$ of $U(1)^4$.

For a one-dimensional $n$-leg register (the coupled register pair in the Entangled QFT case), the gate-manifold products represented by the table are
\begingroup
\allowdisplaybreaks
\begin{align}
  \MM_{\mathrm{QFT}}^{(n)}
    &=U(2)^n\times\bigl(U(1)^4\bigr)^{n(n-1)/2}, \notag\\
  \MM_{\mathrm{TEBD}}^{(n)}
    &=U(2)^n\times U(4)^n, \notag\\
  \MM_{\mathrm{MERA}}^{(n)}
    &=U(2)^n\times U(4)^{2(n-1)}, \notag\\
  \MM_{\mathrm{Rich}}^{(n)}
    &=U(2)^n\times U(4)^{n(n-1)/2}, \notag\\
  \MM_{\mathrm{DCT}}^{(n)}
    &=O(4)^{n(n-1)}
      \times O(2)^{n(n-1)/2+2n} \notag\\*[-2pt]
    &\quad{}\times \bigl(U(1)^4\bigr)^{n-1}, \notag\\
  \MM_{\mathrm{Ent}}^{(m,n)}
    &=\MM_{\mathrm{QFT}}^{(m)}\times\MM_{\mathrm{QFT}}^{(n)} \notag\\*[-2pt]
    &\quad{}\times\bigl(U(1)^4\bigr)^{\min(m,n)}.
  \label{eq:appendix_gate_manifolds}
\end{align}
\endgroup
The exponents count gates, one $U(2)$ factor per leg and one two-qubit factor per bond of the wiring drawn in \cref{fig:topology_circuits} (the $n(n-1)/2$ all-to-all bonds of QFT and RichBasis, the $n$ ring bonds of TEBD, and the $2(n-1)$ hierarchical bonds of MERA), or for the DCT-IV the role counts of \cref{eq:dct4_gate_counts}, so the number of factors is the Gates entry of \cref{tab:circuits}.

The QFT and RichBasis products contain the exact QFT, since each $U^{(4)}$ gate of RichBasis can sit at the corresponding phase gate; Entangled QFT contains the separable QFT at zero coupling, and DCT-IV the bit-reversed exact transform. The ring and hierarchical wirings of TEBD and MERA offer no direct embedding of the QFT gate pattern at the sizes trained here. \Cref{sec:train_robustness} gives each topology's initialization, and \cref{app:training_preservation} shows that every update keeps each gate unitary to machine precision, so the transform stays invertible throughout training; the gates assigned to $O(2)$ and $O(4)$ stay real only to about $10^{-3}$. A 2-D basis uses independent row and column copies apart from Entangled QFT's matched-axis factors, and \cref{tab:circuits} counts every gate of one single-axis circuit, the coupled register pair in the Entangled QFT case.

\subsection{DCT-IV versus DCT-II}
\label{app:dct_suboptimality}

The DCT-II used by JPEG is a strong coding baseline. It approaches the Karhunen--Lo\`eve basis of the AR(1) source of \cref{sec:experiments} as $\rho_{\mathrm{AR}}\to1$ \citep{jain1979sinusoidal, britanak2007} and has classical fast factorizations of its own \citep{chen1977dct, loeffler1989dct, feig1992dct}. The reason not to use it as the recursive anchor is structural.

Using the same reversal $R_{N/2}$ and interleaver $P_N$ as above, the DCT-II radix-2 identity is
\begin{equation}
  \begin{aligned}
  B_N&=\frac{1}{\sqrt2}
    \begin{pmatrix}I_{N/2}&R_{N/2}\\I_{N/2}&-R_{N/2}\end{pmatrix},\\
  C_N^{\mathrm{II}}
  &=P_N\bigl(C_{N/2}^{\mathrm{II}}\oplus C_{N/2}^{\mathrm{IV}}\bigr)B_N.
  \end{aligned}
  \label{eq:dct2_nonuniform}
\end{equation}
\Cref{eq:dct2_nonuniform} has two different children, $C_{N/2}^{\mathrm{II}}$ and $C_{N/2}^{\mathrm{IV}}$. It therefore does not define the one-template self-recursion of \cref{fig:cooley_tukey_to_dct}; an implementation of this recursion must retain two transform templates or add a transform-type leg (an ancilla) to select the child. In contrast, the repeated diagonal block in \cref{eq:dct4_recursion} contains two identical copies of $C_{N/2}^{\mathrm{IV}}$, so no ancilla is required.

\FloatBarrier

\section{Optimization details}
\label{app:riemannian}

\subsection{Riemannian Adam update}
\label{app:radam_update}

This subsection records the implementation-specific update behind
\cref{sec:riemannian}, in the same notation, with $G_t$ a trainable gate at
iteration $t$ and $g_t$ the ambient gradient $\nabla\LL_k$ of the batch-mean
loss of \cref{app:einsum}, evaluated at $G_t$.
For complex gates, the raw JAX gradient is first conjugated to follow the
real-valued Wirtinger convention. The resulting gradient $g_t$ at
$G_t\in U(d)$ is projected as
\begin{equation}
  \begin{aligned}
  \proj_G(g)&=G\,\skew(G^\dagger g),\\
  \skew(A)&=\tfrac12(A-A^\dagger).
  \end{aligned}
  \label{eq:projection}
\end{equation}
For $O(d)$, the skew factor is already real, so the same formula applies; for a phase gate,
$\proj_z(g)=i\,\operatorname{Im}(\bar z\odot g)\odot z$, with $\odot$ the
elementwise product. These projections land
in the tangent space at each point,
\begin{equation*}
  \begin{aligned}
  T_U\,U(d)&=\{UA:A^\dagger=-A\},\\
  T_Q\,O(d)&=\{QA:A^\top=-A\},
  \end{aligned}
\end{equation*}
and $T_z\,U(1)^q=\{i\,\mathbf{t}\odot z:\mathbf{t}\in\RR^q\}$; each tangent
vector is the point itself times a skew factor, which is why the one projection
formula covers the whole family. Writing
$r_t=\proj_{G_t}(g_t)$ and $\widetilde m_{t-1}$ for the transported first
moment, with $\widetilde m_0=v_0=0$ and $t$ counted from $1$, the
elementwise Adam moments are
\begin{align}
  m_t &= \beta_1\widetilde m_{t-1}+(1-\beta_1)r_t,
  &\hat m_t &= \frac{m_t}{1-\beta_1^t}, \\
  v_t &= \beta_2v_{t-1}+(1-\beta_2)|r_t|^2,
  &\hat v_t &= \frac{v_t}{1-\beta_2^t}.
  \label{eq:radam_moments}
\end{align}
Here $v_t$ is a real elementwise accumulator and is not transported; this
elementwise second moment departs from the per-factor scalar of
\citet{becigneul2019riemannian}. With
$(\beta_1,\beta_2,\epsilon)=(0.9,0.999,10^{-8})$, the adaptive increment is
\begin{equation}
  a_t=-\eta\,\hat m_t/(\sqrt{\hat v_t}+\epsilon).
  \label{eq:radam}
\end{equation}
The elementwise division takes $a_t$ off the tangent space; the $\skew$
in \cref{eq:retraction} restores it. In the released benchmark runs of \cref{tab:div2k_repr} and the
perturbation study of \cref{app:exact_disturbance}, the projected
gradient is additionally clipped to a global norm of $1$ before the
moment update, and $\eta$ follows a warmup and cosine-decay schedule;
the robustness studies of \cref{app:direct_training} train at a fixed
step size without clipping.
For unitary gates, the implementation retracts this increment with the
Cayley map quoted in \cref{sec:riemannian},
\begin{equation}
  \begin{aligned}
  W_t&=\skew(a_tG_t^\dagger),\\
  G_{t+1}
    &=\bigl(I-\tfrac12W_t\bigr)^{-1}
      \bigl(I+\tfrac12W_t\bigr)G_t.
  \end{aligned}
  \label{eq:retraction}
\end{equation}
Because $W_t^\dagger=-W_t$, the Cayley factor is unitary. The real analogue
preserves $O(d)$, while a phase gate is updated by normalizing $z_t+a_t$ to
unit modulus. Finally, the first moment is transported by re-projection,
\begin{equation}
  \begin{aligned}
  \widetilde m_t^{(G)}
    &=G_{t+1}\skew(G_{t+1}^\dagger m_t),\\
  \widetilde m_t^{(z)}
    &=i\,\operatorname{Im}(\bar z_{t+1}\odot m_t)\odot z_{t+1}.
  \end{aligned}
\end{equation}
This projection transport is the first-order transport used by the optimizer,
rather than exact geodesic parallel transport.

The released values of this schedule: the benchmark bases of
\cref{tab:div2k_repr} train for $112$ epochs of $50$-image minibatches
($1008$ steps), with $\eta$ warmed up over the first $5\%$ of steps to a
peak of $3 \times 10^{-3}$ and cosine-decayed to $3 \times 10^{-4}$;
per-run values are recorded in each released basis's run-environment file.

\subsection{Constraint preservation}
\label{app:training_preservation}

Let $G_{p,t}\in\mathcal M_p$ be the $p$th of the network's $P'$ gates at
iteration $t$. Projection
changes only the update direction, and factor-wise retraction returns
$G_{p,t+1}\in\mathcal M_p$. Hence every gate remains on its assigned manifold.
In the implementation the $O$-assigned gates are carried by the complex
unitary kernel of \cref{eq:projection}, which keeps \cref{eq:trained_unitarity}
at machine precision at every step; their realness is not separately
re-projected and holds to about $10^{-3}$ in the released bases.
For the ordered composition $\TT(\theta_t)=G_{P',t}\cdots G_{1,t}$,
\begin{equation}
  \TT(\theta_t)^\dagger\TT(\theta_t)
  =G_{1,t}^\dagger\cdots G_{P',t}^\dagger G_{P',t}\cdots G_{1,t}=I.
  \label{eq:trained_unitarity}
\end{equation}
Thus the trained transform remains invertible, and because the gate dimensions
and wiring stay fixed, training changes neither the parameter count nor the
contraction cost of \cref{app:transform_complexity}.

\subsection{Batched evaluation}
\label{app:einsum}

The transform is never materialized as a dense
$2^{m+n}\times 2^{m+n}$ matrix. Instead, $\TT(\theta)\,\mathbf{x}$ is
evaluated on the binary-leg image tensor of \cref{sec:circuits}, with every
gate $G_p$ kept as a separate tensor in the contraction, at the
gate-by-gate cost of \cref{app:transform_complexity}. The gates are
applied in their circuit order, which the topology fixes. This order is not
searched and in particular is not delegated to JAX's default \texttt{einsum}
path policy.
Each gate is an axis-wise contraction against the image tensor: controlled
rotations act on the slice their control selects, and diagonal gates enter as
broadcast multiplications on the labels of their control and target legs,
introducing no new legs. The fixed sequence is compiled once per topology
with JAX \citep{jax2018} and cached. The adjoint $\TT(\theta)^\dagger$ reuses
the same network with conjugated gate tensors and reversed gate legs, so
the forward and inverse transforms share one parameter set.

A minibatch adds one batch leg. The per-image loss $\LL_k$ of
\cref{eq:sparse_basis_problem} is vectorized over the leading batch axis
with the gate tensors broadcast across it, and the training objective is
the batch mean. Reverse-mode differentiation through the compiled
contraction \citep{liao2019differentiable, roavillescas2024probabilistic}
therefore returns one accumulated ambient gradient per gate,
which, once conjugated, is the $g_t$ consumed by the update of
\cref{app:radam_update}. Keeping every gate as a separate tensor in the
contraction preserves its independent manifold constraint; gates that share a
factor manifold are grouped and updated by one stacked
projection--retraction kernel.

\section{Training robustness}
\label{app:direct_training}

\begin{figure}[!htbp]
  \centering
  \includegraphics{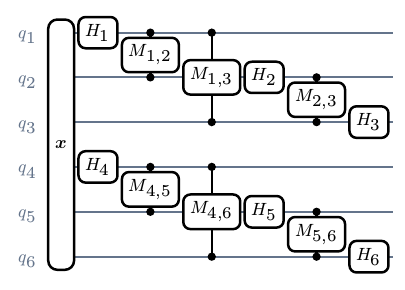}
  \caption{The QFT network of \cref{fig:banner}b with its gates numbered,
  drawn for QFT$(3,3)$. The DIV2K runs use the analogous QFT$(8,8)$ ($72$
  gates). The two registers act on the rows ($q_1$--$q_3$) and columns
  ($q_4$--$q_6$) of the input $\mathbf{x}$}
  \label{fig:app_circuit}
\end{figure}

\begin{figure*}[p]
  \centering
    \includegraphics[width=\textwidth]{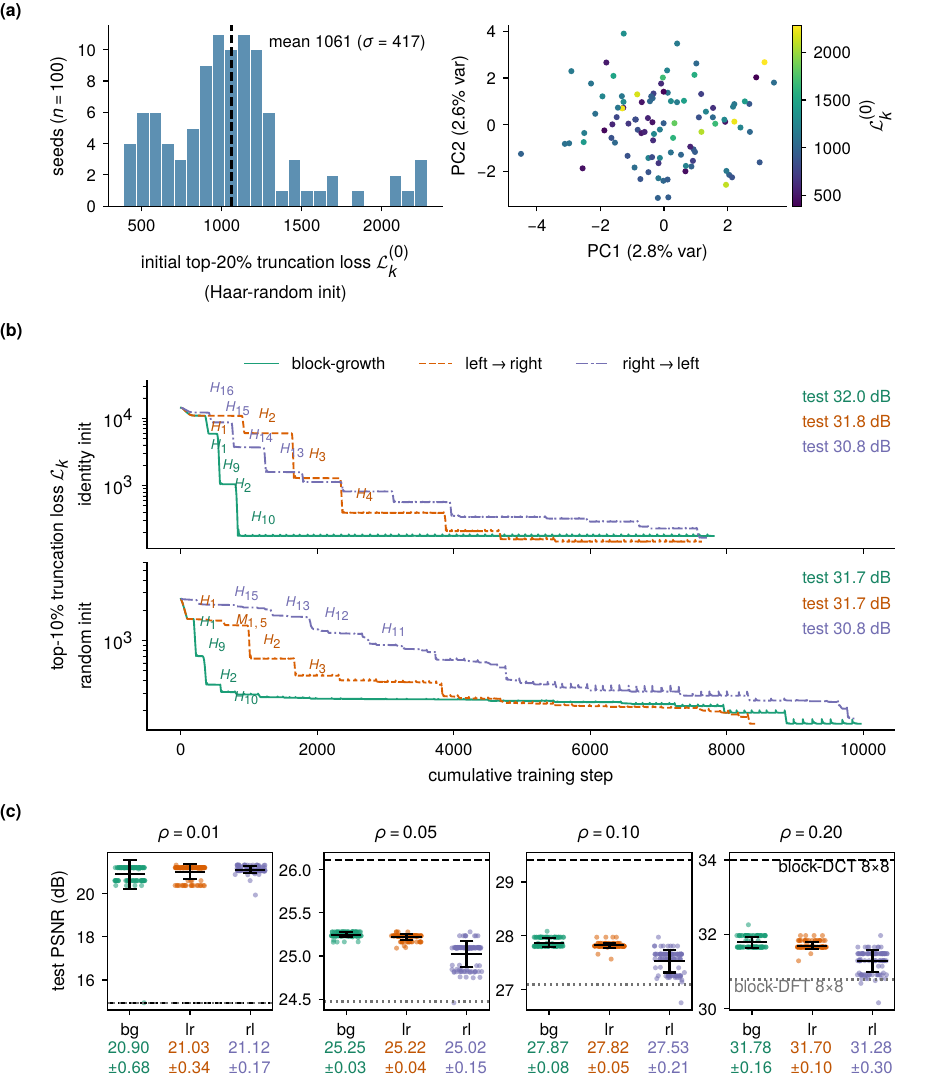}
  {\phantomsubcaption\label{fig:app_seed_robustness_a}}
  {\phantomsubcaption\label{fig:app_unfreeze_dynamics}}
  {\phantomsubcaption\label{fig:app_seed_robustness_c}}
  \caption{Robustness of QFT$(8,8)$ training on DIV2K. (a)~$100$ Haar-random
  initializations: initial top-$20\%$ truncation loss $\LL_k^{(0)}$ (left, mean
  dashed) and a 2-D principal component analysis (PCA) of the init parameters,
  colored by $\LL_k^{(0)}$ (right). (b)~Unfreeze dynamics: top-$10\%$ truncation
  loss $\LL_k$ per thaw ordering from identity (top) and Haar-random (bottom)
  init. Labels mark the gates behind the largest drops, and endpoints give the
  test PSNR at $\rho = 0.20$. (c)~Per-seed test PSNR for the block-growth (bg),
  left-to-right (lr) and right-to-left (rl) orderings, $100$ Haar-random
  initializations each under the top-$20\%$ objective, with mean$\,\pm\,\sigma$
  as error bars and numbers. Dashed and dotted lines are the $8 \times 8$ block
  DCT and DFT. At $\rho = 0.20$, $297/300$ runs beat the block DFT (worst seed
  $30.16$~dB)}
  \label{fig:app_direct}
\end{figure*}

This appendix assesses the robustness of the training procedure for the
$N=2^8$ QFT on the DIV2K dataset, using the gate numbering of
\cref{fig:app_circuit}. \Cref{app:unfreeze} varies the order in which the
gates are trained and their initialization, and a seed-robustness study varies
the random seed behind the initialization and the batch subsample.
\Cref{fig:app_direct} collects the results. Test PSNRs in this appendix are
scored on the original $50$-image test set, as in
\cref{app:exact_disturbance}. The full training traces of the unfreeze runs,
and the trained basis and endpoint scores of each reseeded run, are included
in the benchmark archive \citep{an2026benchmarks}.

\subsection{Gate unfreezing}
\label{app:unfreeze}

The trained operator is two independent copies of the QFT network drawn in
\cref{fig:banner}b, one per image axis. \Cref{fig:app_circuit} draws the
same network for three qubits per axis, with its gates numbered. $H_i$ is the Hadamard-role $U(2)$ gate on
qubit $i$ and $M_{i,j}$ ($i<j$) the controlled phase coupling qubits $i$ and
$j$ within an axis, $72$ gates in total at $m=n=8$. Wires are numbered $q_1$
to $q_{16}$ across the two registers, matching the gate numbering of the
released runs rather than the per-register bit labels $q_0,\dots,q_{n-1}$ of
\cref{app:decompose}, with the axis-1 register carrying $q_9$ to $q_{16}$.

We thaw the gates one at a time: each stage thaws one gate and
trains all unfrozen gates to a plateau (Riemannian grad-norm $\lVert g\rVert < 10^{-5}$ or
$\lvert\Delta \LL_k\rvert < 10^{-5}$, minimum $5$ steps, $800$-step cap) on a fixed
$50$-image batch under the top-$10\%$ truncation loss $\LL_k$, with the Adam
moments reset between stages. We compare three thaw
orderings and two initializations: the QFT-family identity initialization
$\TT(\theta_{\mathrm{id}})\,\mathbf{x}=\mathbf{x}$, and
Haar-random, which draws each Hadamard-role gate from the Haar distribution on
$U(2)$, the unique unitarily invariant (uniform) measure on the
group \citep{mezzadri2007random}, and each controlled phase uniformly, with the
seed shared across orderings.

Written on these gates, the three orderings are:
\begin{itemize}
  \item \emph{Block-growth} (\texttt{bg}): at stage $s$, thaw, on each axis in
  turn, qubit $s$ and then its couplings to earlier qubits, gate by gate,
  growing a square
  $2^{s}\times2^{s}$ transform; long-range couplings thaw only in the late stages.
  \item \emph{Left-to-right} (\texttt{lr}): the QFT construction order (each
  qubit's Hadamard-role gate, then its couplings), axis 0 before axis 1.
  \item \emph{Right-to-left} (\texttt{rl}): the exact reverse.
\end{itemize}
The \texttt{bg} partner of $H_1$ is therefore $H_9$, which is why $H_9$ is
among the earliest large drops in \cref{fig:app_unfreeze_dynamics}.

\begin{figure*}[!t]
\centering
\includegraphics[width=\textwidth]{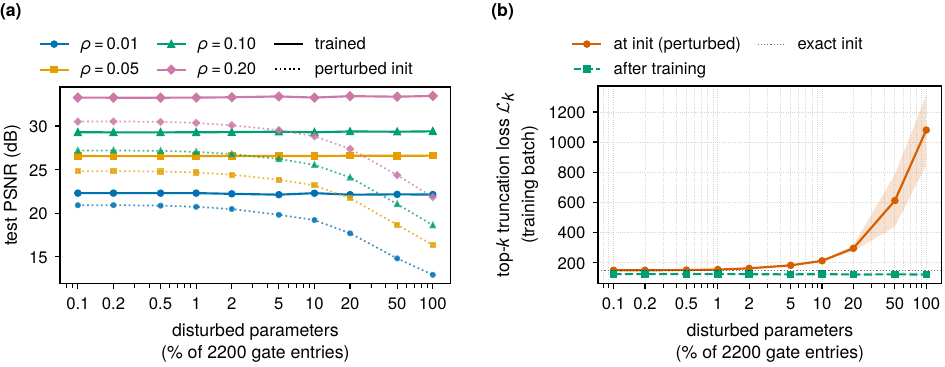}
{\phantomsubcaption\label{fig:disturbance_recovery}}
{\phantomsubcaption\label{fig:disturbance_loss}}
\caption{Disturbance sweep of the exact DCT-IV initialization on DIV2K-8q at noise
scale $\sigma_{\mathrm{jit}} = 0.1$, against the disturbed fraction $f$ of the
$2200$ gate entries (means over three seeds). (a)~The perturbed-init PSNR
(dotted) falls with $f$, while the trained PSNR (solid) stays within
${\sim}0.2$~dB of the undisturbed reference. (b)~Training-batch loss at the
perturbed init and after $1010$ steps, with the exact-init loss dotted and
$\pm1$ standard deviation shaded}
\label{fig:disturbance}
\end{figure*}

\Cref{fig:app_unfreeze_dynamics} shows the resulting staircase of the truncation loss $\LL_k$.
The ordering changes the trajectory, \texttt{bg} collapsing within a few
stages and \texttt{rl} descending late in long steps, but all six
order\,$\times$\,init combinations end within $1.2$~dB of one another, with
the two \texttt{rl} runs trailing by ${\sim}1$~dB, and every stage terminates
on the loss-$\Delta$ plateau rather than the step cap. Training loss and test PSNR
order the runs differently, since the loss is a top-$10\%$ proxy on the
training batch while the PSNR is a $20\%$-keep score on the held-out set. The
identity-init \texttt{bg} run, for instance, ends highest in training loss yet best in
test PSNR.

\section{Initialization noise}
\label{app:exact_disturbance}

The controlled DCT-IV circuit of \cref{app:dct4_circuit} is initialized at the
exact analytic transform, and \cref{tab:div2k_repr} reports its relaxed optimum
after training. A disturbance sweep tests whether progressively broader local
perturbations destroy the useful structure of the untrained transform, and
whether training recovers from the perturbed starts. Starting from the exact init, we
select a random fraction $f$ of the DCT-IV's stored gate entries (counted below),
apply on-manifold Gaussian perturbations of fixed scale $\sigma_{\mathrm{jit}}=0.1$, and retrain
for $1010$ steps ($101$ epochs of ten $50$-image minibatches) of the top-$10\%$ truncation loss $\LL_k$ on the $500$-image DIV2K-8q
pool, scoring test PSNR at four keep ratios $\rho$ on the original $50$-image
test set (the first half of the expanded set behind \cref{tab:div2k_repr}, so
PSNRs differ slightly from its means). Ten disturbance rates from $f = 0.1\%$ ($2$ selected entries) to
$100\%$ (all of them), three seeds each, vary the \emph{breadth} of an
on-manifold disturbance at the fixed local scale $\sigma_{\mathrm{jit}}$; robustness to
larger-magnitude disturbances is not probed.

\paragraph{On-manifold perturbation.}
Evaluating \cref{eq:dct4_gate_counts} at $n=8$ for the two registers gives
$214$ gates: $112$ two-leg $X$ gates and $102$ structured
single-qubit gates. The implementation stores the $X$ gates as $4\times4$
arrays and the others as $2\times2$ arrays, giving
$112\cdot16+102\cdot4=2200$ real-valued entries at exact initialization. The jitter treats these entries as one flat vector of length
$E=2200$ but projects each affected gate separately, so each perturbed start
remains a valid unitary transform with the DCT-IV wiring. It is no longer the
exact DCT-IV. We draw a
uniform subset $S \subseteq \{1,\dots,E\}$ with $|S| = \lfloor fE \rceil$. For
each gate $G$ it touches, we add Gaussian noise to the selected entries and
re-project onto the gate's manifold,
\begin{equation}
\hat{G} = G + \sigma_{\mathrm{jit}}\,(\Xi \odot B_G), \qquad G' = \operatorname{polar}(\hat{G}),
\label{eq:disturbance}
\end{equation}
where $\Xi_{ij}\sim\mathcal{N}(0,1)$, $B_G$ masks $G$'s selected
entries, and $\operatorname{polar}(\hat{G})$, the orthogonal factor of the polar
decomposition, is the nearest orthogonal matrix to $\hat{G}$ (the controlled
sign gate, stored as the phase stub $D(0,0,0,\pi)$ of \cref{tab:gate_relaxations},
is jittered in its phase instead). A gate moves
only if one of its entries is selected. Because the polar projection acts on the
whole gate, $f$ specifies how broadly gates are seeded with noise rather than the
final fraction of entries that differ.

\paragraph{Training recovers the tested perturbations.}
For the tested noise scale and seeds, the seed-averaged trained endpoint is
insensitive to $f$ (solid lines in \cref{fig:disturbance_recovery}), staying
within ${\sim}0.2$~dB of the undisturbed exact-init endpoint ($33.25$~dB at
$\rho = 0.20$) at every keep ratio, even though the perturbed-init PSNR falls
by up to $8.6$~dB across the sweep (dotted lines).

The same recovery shows up in the training loss (\cref{fig:disturbance_loss}). The
perturbed init's truncation loss $\LL_k$ on a fixed $50$-image training batch, the value the optimizer
sees at step $0$, rises from $148$ at the exact init to $1080$ at $f=100\%$.
After $1010$ steps every perturbed init returns to essentially the exact-init
training loss ($121$ at $f=100\%$ versus $123$ at $f=0$).

\end{appendices}

\bibliography{references}

\end{document}